\documentclass[twocolumn]{aastex63}
\usepackage{amsmath}

\newcommand{\afluxa}{450~$\mu$m}
\newcommand{\afluxb}{850~$\mu$m}
\newcommand{\afluxc}{870~$\mu$m}

\newcommand{\fafluxa}{$f_{450\,\mu{\rm m}}$}
\newcommand{\fafluxc}{$f_{870\,\mu{\rm m}}$}
\newcommand{\fafluxd}{$f_{2\,{\rm mm}}$}

\newcommand{\mma}{$f_{2\,{\rm mm}}/f_{450\,\mu{\rm m}}$}
\newcommand{\mmc}{$f_{2\,{\rm mm}}/f_{870\,\mu{\rm m}}$}

\newcommand{\amm}{$f_{450\,\mu{\rm m}}/f_{2\,{\rm mm}}$}
\newcommand{\bmm}{$f_{450\,\mu{\rm m}}/f_{870\,\mu{\rm m}}$}
\newcommand{\bbmm}{$f_{450\,\mu{\rm m}}/f_{850\,\mu{\rm m}}$}
\newcommand{\cmm}{$f_{870\,\mu{\rm m}}/f_{2\,{\rm mm}}$}
\newcommand{\ccmm}{$f_{850\,\mu{\rm m}}/f_{2\,{\rm mm}}$}

\newcommand{\btradmm}{$f_{870~\mu{\rm m}}/f_{450~\mu{\rm m}}$}

\shortauthors{Cowie et al.}

\begin{document}

\title{2~mm Observations  and the Search for High-Redshift Dusty Star-forming Galaxies}

\correspondingauthor{Lennox Cowie}
\email{cowie@ifa.hawaii.edu}

\author[0000-0002-6319-1575]{L.~L.~Cowie}
\affiliation{Institute for Astronomy, University of Hawaii,
2680 Woodlawn Drive, Honolulu, HI 96822, USA}

\author[0000-0002-3306-1606]{A.~J.~Barger}
\affiliation{Department of Astronomy, University of Wisconsin-Madison,
475 N. Charter Street, Madison, WI 53706, USA}
\affiliation{Department of Physics and Astronomy, University of Hawaii,
2505 Correa Road, Honolulu, HI 96822, USA}
\affiliation{Institute for Astronomy, University of Hawaii, 2680 Woodlawn Drive,
Honolulu, HI 96822, USA}

\author[0000-0002-8686-8737]{F.~E.~Bauer}
\affiliation{Instituto de Astrof\'isica and Centro de Astroingenier\'ia, Facultad de F\'isica, 
Pontificia Universidad Cat\'olica de Chile, Casilla 306, Santiago 22, Chile}
\affiliation{Millennium Institute of Astrophysics (MAS), Nuncio Monse{\~{n}}or S{\'{o}}tero 
Sanz 100, Providencia, Santiago, Chile} 
\affiliation{Space Science Institute,
4750 Walnut Street, Suite 205, Boulder, Colorado 80301, USA}

\begin{abstract}
Finding high-redshift ($z\gg4$) dusty star-forming galaxies 
is extremely challenging. It has recently been suggested that
millimeter selections may be the best approach, since the negative K-correction makes galaxies at a given 
far-infrared (FIR) luminosity brighter at $z\gtrsim4$ than those at $z=2$--3.
Here we analyze this issue using  a deep ALMA 2~mm sample obtained by targeting
ALMA \afluxc\ priors (these priors were the result of targeting SCUBA-2 \afluxb\ sources) in the GOODS-S.
We construct prior-based 2~mm galaxy number counts and compare them with 
published blank field-based 2~mm counts, finding good agreement down to 0.2~mJy.
Only a fraction of the current 2~mm extragalactic background light 
is resolved, and we estimate what observational depths may be needed to resolve it fully.
By complementing the 2~mm  ALMA data with a deep SCUBA-2 \afluxa\ sample, we exploit
the steep gradient with redshift of the 2~mm to \afluxa\ flux density ratio to
estimate redshifts for those galaxies without spectroscopic or robust optical/near-infrared photometric redshifts.
Our observations measure galaxies with star formation rates in excess of
250~M$_\odot$~yr$^{-1}$. For these galaxies, the star formation rate densities
fall by a factor of 9 from $z=2$--3 to $z=5$--6.
\end{abstract}


\section{Introduction}

\begin{figure*}
\includegraphics[width=9cm,angle=0]{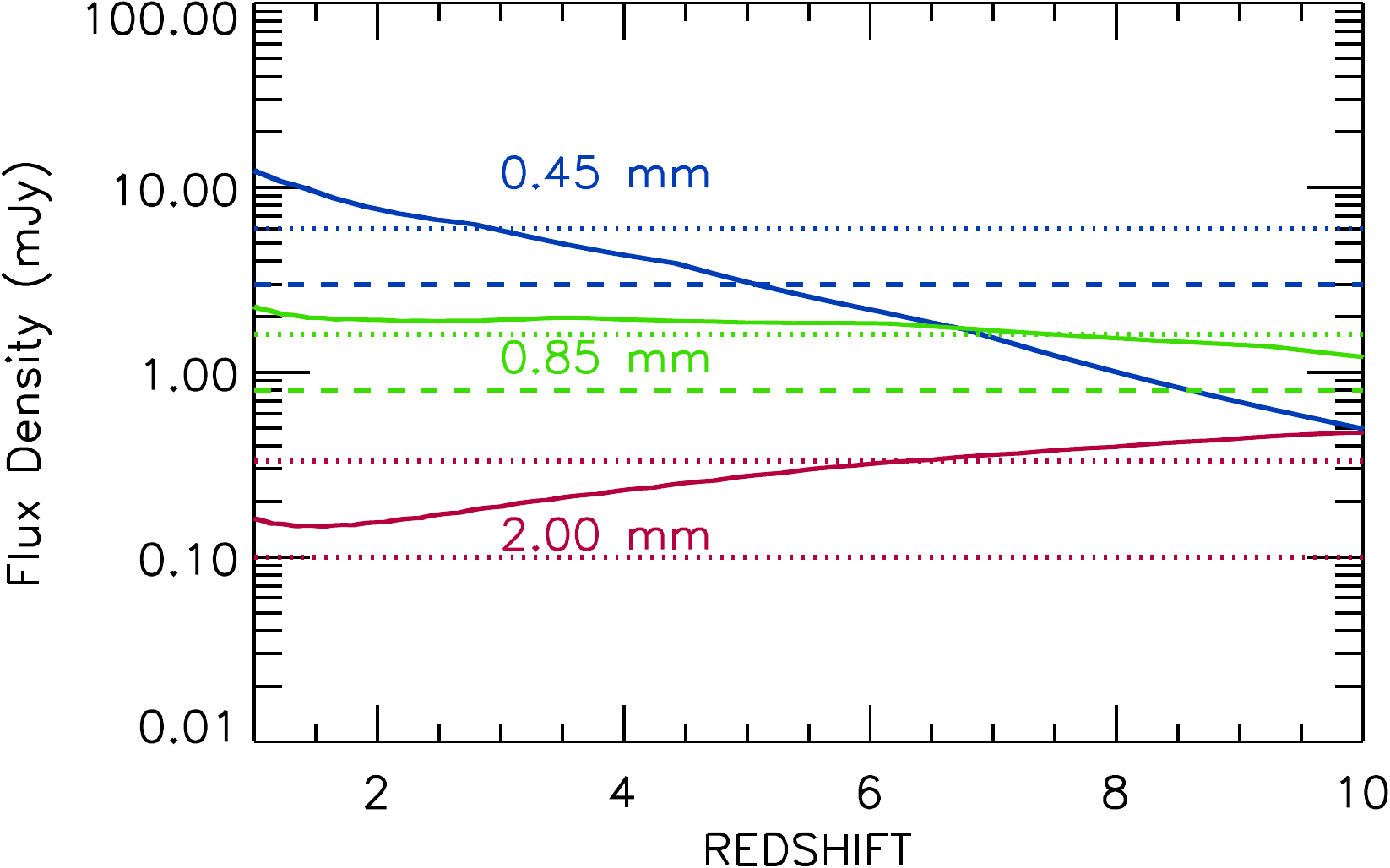}
\includegraphics[width=9cm,angle=0]{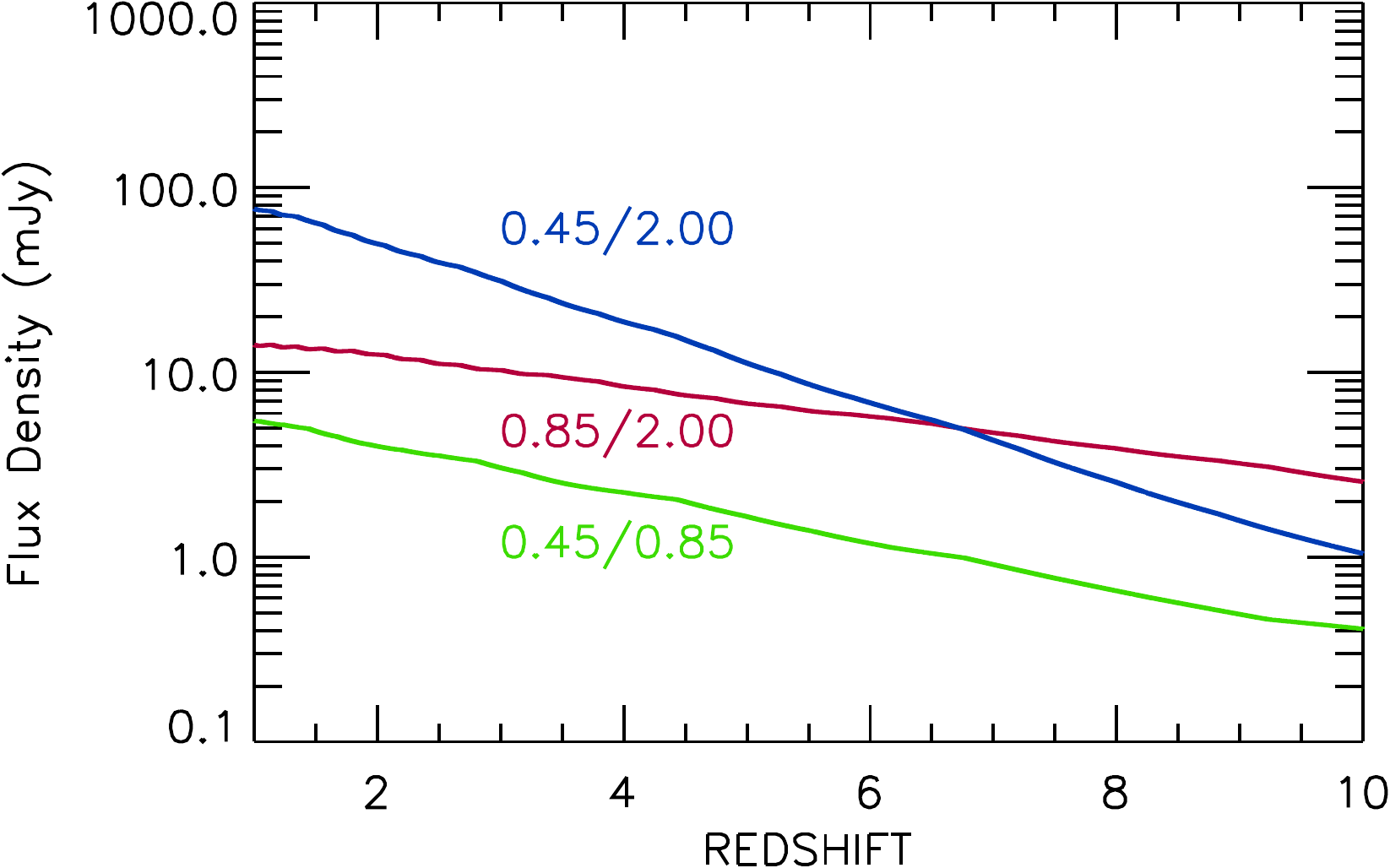}
\caption{
(a) 
Flux vs. redshift curves (solid) for a source with
$L_{\rm FIR}=10^{12}$\,L$_\odot$ at \afluxa\ (blue), \afluxb\ (green),
and 2~mm (red), using Arp~220 as a template for the SED shape \citep{silva98}.
The \afluxa\ flux has the steepest gradient and hence is critical for identifying
high-redshift candidates.
From top to bottom, the $4\sigma$ flux limits are 
6~mJy at \afluxa\ for SCUBA-2 blank fields \citep{barger22} (blue dotted)
and a factor of 2 deeper for SCUBA-2 lensing cluster fields \citep{cowie22} (blue dashed),
assuming a factor of 2 magnification typical of the fainter sources in
\citet{cowie22}; the 1.65~mJy confusion limit at \afluxb\
for SCUBA-2 blank fields \citep{cowie17} (green dotted) and again a factor of 2 deeper for 
SCUBA-2 lensing cluster fields (green dashed); 
0.33~mJy for the ALMA MORA 2~mm survey \citep{casey21} (red dotted); and
0.13~mJy for the ALMACAL 2~mm survey \citep{chen23} (red dotted).
(b) 
Flux ratio vs. redshift curves for \amm\ (blue), \ccmm\ (red), and \bbmm\ (green).
}
\label{figselect}
\end{figure*}

The discovery of the far-infrared (FIR) extragalactic background light (EBL) by COBE 
demonstrated that about half of the universe's starlight at UV/optical wavelengths is 
absorbed by dust and reradiated into the FIR \citep{puget96,fixsen98}.
Moreover, from individual source measurements in the $z$ = 1--4 redshift range, it has been 
found that up to five times as much starlight is radiated into 
the FIR as is seen in the UV/optical \citep[e.g.,][]{wang06,zavala21}. 
We therefore need to study both the unobscured and dust-obscured populations of 
galaxies across cosmic time to obtain a complete picture of the star formation in our universe.
However, even multiwavelength galaxy number counts alone---the projected galaxy surface 
density with flux---can provide critical constraints on galaxy modeling 
and help us to understand the physical processes behind galaxy formation and evolution
\citep[e.g.,][]{shimizu12,schaye15,dave19,lagos20,popping20}.

The galaxy number counts are now well established at both 
\afluxa\ and \afluxb\ \citep[e.g.,][]{chen13b,hsu16,wang17,zavala17,bethermin20,barger22}.
Although there have been efforts to measure the 2~mm number counts with
GISMO \citep{staguhn14,magnelli19} on the Institute for Radioastronomy at Millimeter 
Wavelengths (IRAM)~30~m telescope and with the Atacama Large 
Millimeter/submillimeter Array (ALMA) \citep{zavala21,chen23}, 
these map only a fraction of the EBL. With NIKA2 \citep{adam18} now on IRAM and
the commissioning of TolTEC \citep{wilson20} on the Large Millimeter Telescope~50~m, 
which has a higher resolution (at 2~mm, TolTEC's resolution is $9\farcs5$  versus NIKA2's $17\farcs5$),
it is of interest to determine how deep of observations will be needed to measure the full EBL at 2~mm.

Despite several decades of work, determining the number of dusty star-forming galaxies (DSFGs) 
at high redshifts continues to be a major challenge, since high-redshift DSFGs are often too faint
for optical/near-infrared (NIR) spectroscopic redshifts
\citep[e.g.,][]{cowie09,wang09,barger14,dud20,smail21}.
It has been suggested \citep[e.g.,][]{casey21} that
observations at 2~mm might provide a promising means for finding 
such galaxies, since for a given FIR luminosity, the negative K-correction makes galaxies
at $z\gtrsim4$ slightly brighter at 2~mm than those at $z\sim2$--3 (see Figure~\ref{figselect}(a));
however, the curve is quite flat.
By comparison, the \afluxb\ flux drops by these higher redshifts.
Then, through the use of the 2~mm to \afluxc\ flux ratio, one can try to estimate redshifts for these galaxies
\citep[e.g.,][]{casey21,cooper22}. However, as we discuss in this paper, redshifts are better estimated using 
the flux ratio of 2~mm to a shorter bandpass, such as \afluxa, where the gradient is steeper
(see Figure~\ref{figselect}(b)). Thus, observations at shorter wavelengths are critical for 
separating high-redshift candidates from lower redshift sources.

DSFGs are sparse (e.g., $\sim1$ source per 1~arcmin$^2$ at an \afluxb\ flux of 2~mJy; 
\citealt{hsu16}), 
so it is very inefficient to map them with small field-of-view instruments, though the ALMA 
sensitivity is such that modest samples can be generated with enough invested time 
\citep[e.g.,][]{dunlop17,gonzalez17cont,franco18,hatsukade18,aravena20,casey21,gomez22}. 
Instead, the best 
method to select high-redshift DSFGs systematically remains submillimeter/millimeter
surveys from large-aperture, ground-based telescopes, as these provide the fields-of-view 
necessary to detect significant samples. These samples can then be efficiently followed up 
with interferometric observations 
\citep[e.g.,][]{barger12,hodge13,cowie17,cowie18,cowie22,stach19,cooper22}.
In fact, using single-dish priors can provide an enormous gain in speed over direct interferometric 
searches. At 2~mJy, there are about 3000 \afluxb\ sources
per 1~deg$^2$ \citep[e.g.,][]{hsu16}. This is the
number of targeted ALMA pointings one would need
to image this population.  In contrast, given
the ALMA $16\farcs9$ FWHM at this wavelength, one would
need about 57000 ALMA pointings to image fully
this area at the same level. Thus, targeted ALMA imaging 
gives a speed gain of a factor of roughly 19 over ALMA mosaicking.

In this paper, we present new ALMA 2~mm observations of the ALMA \afluxc\ GOODS-S 
sample from \citet{cowie18}, which was based on ALMA
follow-up of the confusion-limited SCUBA-2 \afluxb\ observations of the field.
We construct the 2~mm cumulative number counts, which we compare with the literature,
and we estimate how deep of 2~mm observations are needed to fully resolve the EBL
at this wavelength.
In combination with new deep SCUBA-2 \afluxa\ observations of the field, we 
demonstrate the advantages of using the 2~mm to \afluxa\ flux ratio for identifying
candidate high-redshift DSFGs.

In Section~\ref{secdata}, we describe our new ALMA 2~mm and
SCUBA-2 \afluxa\ observations and data reduction. 
In Section~\ref{secdep}, we examine the dependence of the 2~mm to \afluxc\ flux density 
ratio on the \afluxc\ flux density.
In Section~\ref{secnumcts}, we construct 2~mm cumulative 
and differential number counts, which we compare with the literature.
In Section~\ref{sechighz}, we consider three flux density ratios,
\amm, \cmm, and \bmm,
for identifying candidate high-redshift DSFGs. 
We then estimate the star formation history.
In Section~\ref{secsum}, we summarize our results.

We assume $\Omega_M=0.3$, $\Omega_\Lambda=0.7$, and
$H_0=70$~km~s$^{-1}$~Mpc$^{-1}$ throughout.

\begin{figure}
\includegraphics[width=8.5cm,angle=0]{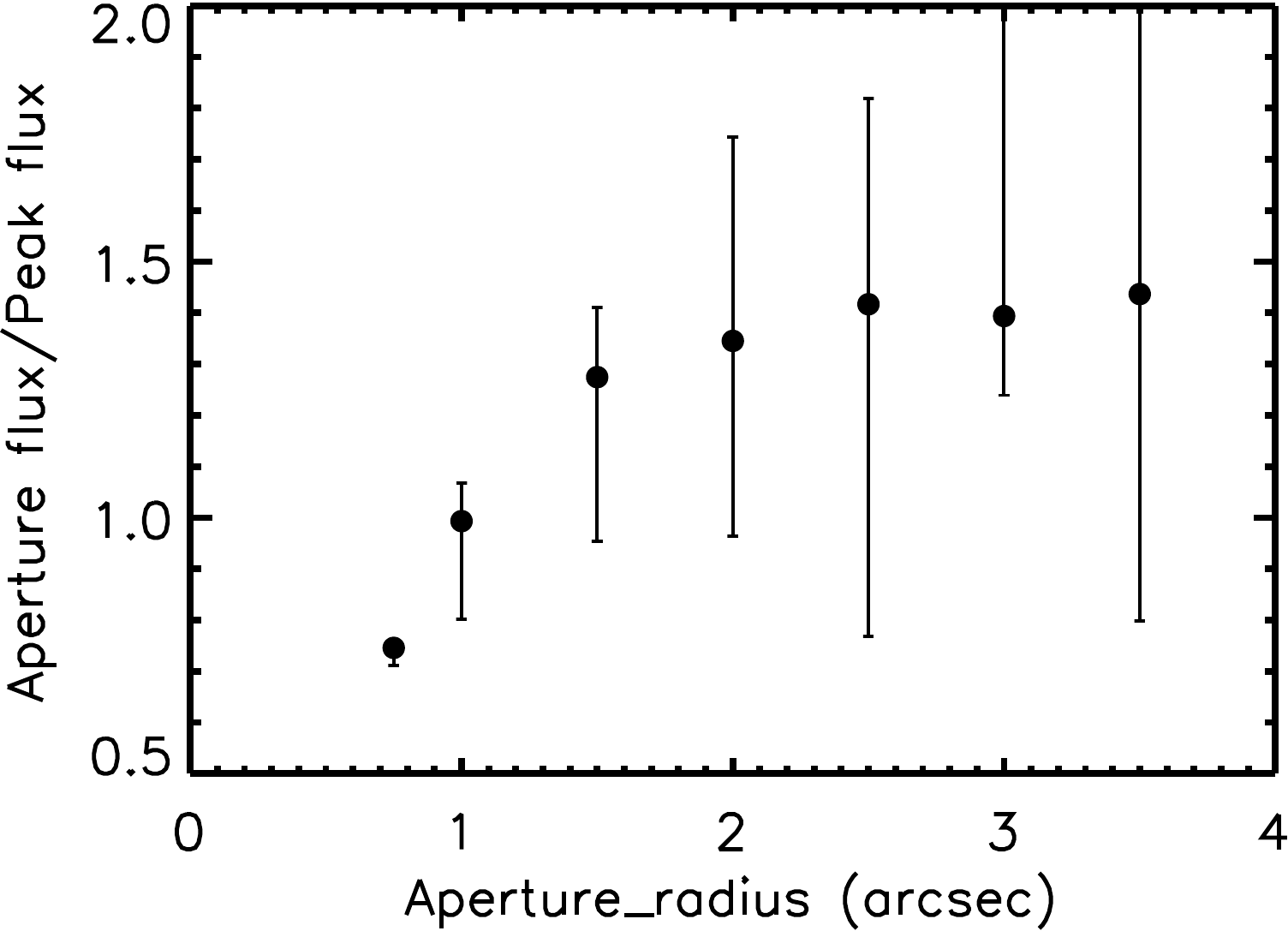}
\caption{
Median ratio of the aperture flux to the peak flux vs. the aperture radius for the 37 ALMA sources
with 2~mm peak flux densities $>0.1$~mJy. 
The error bars show the 68\% confidence range on the median.
}
\label{apcor}
\end{figure}

\section{Data}
\label{secdata}

\subsection{ALMA Observations}
\label{almaobs}
\citet{cowie18} provided a catalog of $>4\sigma$ sources 
from a SCUBA-2 \afluxb\ survey of the GOODS-S.
They also presented the 75 \afluxc\ ALMA sources ($>4.5\sigma$) 
(their Table~4) that resulted from high-resolution interferometric follow-up 
observations of this sample.
In ALMA program \#2021.1.00024.S (PI: F. Bauer),
we made ALMA spectral linescans in band~6 (central wavelength of 1.24~mm),
band~4 (1.98~mm; hereafter, 2~mm), and band~3 (3.07~mm) of 57 sources 
in this sample. We focused on those sources with
\afluxc\ flux densities above 1.8~mJy and without a very well-determined spectroscopic redshift,
since spectroscopic redshifts were the primary goal of the observations.

In McKay et al.\ (2023), we presented the full ALMA data set (their Table~1) and
fit the FIR spectral energy distributions (SEDs) of the sources to constrain the emissivity
spectral indices and effective dust temperatures. 
Here we focus on the band~4 data, since there has been relatively little analysis of
2~mm number counts and faint 2~mm source properties in the literature.

In Table~\ref{tabCAT}, we give our ALMA 2~mm flux densities in a table of \citet{cowie18}'s
75 sources to show which of the sources we observed.
Of the 50 sources with 870~$\mu$m flux densities above 1.8~mJy and lying within a radial offset of 
$5\farcm65$ from the SCUBA-2 image center
(mean aim-point in J2000.0 coordinates: R.A.~$03$:32:26.49, decl.~$-27$:48:29.0),
there are only 5 sources that we did not observe in band~4 (4 because they already had good 
spectroscopic redshifts; the fifth was inadvertently omitted).

The ALMA data were downloaded and calibrated using {\sc casa} version 6.2.1-7 
and PI scripts provided by ALMA. 
We visually inspected various diagnostic plots associated with the calibration 
to confirm that there were no unusual problems with particular antennas or baselines.
The visibilities from individual spectral setups
across each band were aggregated using \texttt{concat}, 
and dirty continuum images were generated 
using \texttt{tclean}, adopting 0\farcs25 pixels, natural weighting,
and a ``common'' restoring beam.
Based on the rms noise from the dirty images, cleaned continuum images were 
generated by adopting multi-threshold auto-masking with default values, assuming 10000 
clean iterations, a flux threshold set to 3 times the rms ($\approx$0.06 mJy), pixel scales 
of 0, 5, and 10, and robust=0.5. As none of our targets are particularly bright, the dirty 
and clean images are nearly identical in terms of peak flux densities, noise, etc. 
We visually inspected the image products, confirming that there were no unusual noise 
residuals or patterns in the images.
The resulting band~4 images had a central frequency of 151.188~GHz,
a bandwidth of 23.375~GHz, and a beam of 
$\theta_{\rm beam}{=}$ 1\farcs21$\times$ 1\farcs10.

We additionally 
generated full-band spectral cubes with the same \textit{tclean} parameters, 
adopting a native resolution of 16~MHz.

The peak flux densities, even in the tapered images, slightly underestimate the total flux densities
\citep[e.g.,][]{cowie18}.
The reason for this is that the sources are resolved. There are two methods we can use to
estimate the total flux densities: We can take
the ratio of the aperture measurements made in a
range of aperture radii to the peak measurements, or we can fit
the sources in the uv plane \citep[e.g.,][]{bethermin20}. Both give similar
corrections \citep{cowie18}, and here we adopt the simpler
aperture method. Because the aperture fits are noisy, we use
a single average correction for all the sources.
As we show in Figure~\ref{apcor}, the median multiplicative correction asymptotes beyond
an aperture radius of $\approx1\farcs5$, and we adopt this as our preferred radius,
giving a multiplicative correction of 1.3.

\startlongtable
\begin{deluxetable*}{cccccccrccc}
\tabletypesize{\scriptsize}
\tablewidth{0pt}
\setcounter{table}{0}
\tablecaption{Flux and Redshift Table \label{tabCAT}}
\tablehead{ \\
C18 & & & R.A. & decl. & \multicolumn{2}{c}{Total} & \multicolumn{2}{c}{Peak} & \multicolumn{2}{c}{Peak} \\
No. & Redshift & Ref. & & & $f_{870\,\mu{\rm m}}$ &  Error & 
$f_{450\,\mu{\rm m}}$ & Error & $f_{2\,{\rm mm}}$ & Error \\
& & & (J2000) & (J2000) & \multicolumn{2}{c}{(mJy)} & \multicolumn{2}{c}{(mJy)}
& \multicolumn{2}{c}{(mJy)}  \\ 
(1) & (2) & (3) & (4) & (5) & (6) & (7) & (8) & (9) & (10) & (11)}
\startdata
1 & 2.574 &   10 &   53.030373 &   -27.855804 &   8.93 &   0.21 &   22.99 &   3.46    &   0.61 &   0.02 \cr 
2 & 3.690 &   10 &   53.047211 &   -27.870001 &   8.83 &   0.26 &   14.88 &   3.26    &   0.54 &   0.02 \cr 
3 & 2.648 &   10 &   53.063877 &   -27.843779 &   6.61 &   0.16 &   18.68 &   2.46    &   0.33 &   0.02 \cr 
4 & 2.252 &   4 &   53.020374 &   -27.779917 &   6.45 &   0.41 &   25.81 &   3.39    &   0.27 &   0.04 \cr 
5 & 2.309 &   7 &   53.118790 &   -27.782888 &   6.39 &   0.16 &   13.67 &   1.87    &   0.26 &   0.03 \cr 
6 &   \nodata  &    \nodata &   53.195126 &   -27.855804 &   5.90 &   0.18 &   13.96 &   3.37    &   0.41 &   0.03 \cr 
7 & 3.672 &   10 &   53.158371 &   -27.733612 &   5.60 &   0.14 &   5.49 &   2.78    &   0.38 &   0.02 \cr 
8 & 2.69 (2.62--2.73)  &   0 &   53.105247 &   -27.875195 &   5.18 &   0.22 &   16.54 &   2.76    &   0.25 &   0.02 \cr 
9 & 2.322 &   10 &   53.148876 &   -27.821167 &   5.09 &   0.12 &   17.27 &   2.21    &   0.32 &   0.04 \cr 
10 & 2.41 (2.34--2.46)  &   0 &   53.082085 &   -27.767279 &   4.90 &   0.29 &   17.39 &   2.27    &   0.26 &   0.02 \cr 
11 &   \nodata  &    \nodata &   53.079376 &   -27.870806 &   4.76 &   0.29 &   14.21 &   2.79    &   0.28 &   0.02 \cr 
12 & 3.764 &   1 &   53.142792 &   -27.827888 &   4.73 &   0.16 &   13.09 &   2.17    &   0.00 &   0.00 \cr 
13 & [2.73 (2.68--4.43)]  &   0 &   53.074837 &   -27.875916 &   4.69 &   0.81 &   7.04 &   2.84    &   0.33 &   0.02 \cr 
14 & 2.73 (2.63--2.80)  &   0 &   53.092335 &   -27.826834 &   4.64 &   0.17 &   17.19 &   1.86    &   0.18 &   0.02 \cr 
15 & 2.14 (2.06--2.23)  &   0 &   53.024292 &   -27.805695 &   3.93 &   0.15 &   11.70 &   3.23    &   0.17 &   0.02 \cr 
16 & 3.37 (3.17--3.41)  &   0 &   53.082752 &   -27.866585 &   4.31 &   0.15 &   9.10 &   2.78    &   0.28 &   0.03 \cr 
17 & 3.57 (3.45--3.81)  &   0 &   53.146629 &   -27.871029 &   3.80 &   0.18 &   9.26 &   2.56    &   0.21 &   0.02 \cr 
18 & 3.847 &   8 &   53.092834 &   -27.801332 &   5.21 &   0.32 &   12.66 &   1.78    &   0.23 &   0.02 \cr 
19 & [4.47 (4.07--5.65)]  &   0 &   53.108795 &   -27.869028 &   3.62 &   0.17 &   7.34 &   2.80    &   0.24 &   0.02 \cr 
20 & 1.93 (1.89--1.96)  &   0 &   53.198292 &   -27.747889 &   3.61 &   0.30 &   8.46 &   3.68    &   0.16 &   0.02 \cr 
21 & 3.78 (3.70--3.83)  &   0 &   53.178333 &   -27.870222 &   3.55 &   0.20 &   8.90 &   3.09    &   0.18 &   0.02 \cr 
22 & 2.698 &   9 &   53.183460 &   -27.776638 &   3.38 &   0.32 &   12.05 &   2.46    &   0.00 &   0.00 \cr 
23 & 1.58 (1.56--1.61)  &   0 &   53.157207 &   -27.833500 &   3.32 &   0.29 &   8.58 &   2.42    &   0.14 &   0.02 \cr 
24 & 1.96 (1.90--1.99)  &   0 &   53.102791 &   -27.892860 &   3.25 &   0.14 &   8.65 &   2.87    &   0.18 &   0.02 \cr 
25 & 2.696 &   9 &   53.181377 &   -27.777557 &   3.18 &   0.23 &   9.41 &   2.42    &   0.00 &   0.00 \cr 
26 & 3.78 (3.70--3.82)  &   0 &   53.070251 &   -27.845612 &   3.15 &   0.25 &   3.06 &   2.46    &   0.19 &   0.02 \cr 
27 & [1.78 (1.67--2.73)]  &   0 &   53.014584 &   -27.844389 &   3.05 &   0.19 &   9.49 &   4.43    &   0.16 &   0.03 \cr 
28 & [3.33 (3.20--4.38)]  &   0 &   53.139290 &   -27.890722 &   2.89 &   0.37 &   4.63 &   2.89    &   0.13 &   0.02 \cr 
29 &   \nodata  &    \nodata &   53.137127 &   -27.761389 &   2.82 &   0.28 &   15.27 &   2.41    &   0.13 &   0.02 \cr 
30 & 1.86 (1.81--1.92)  &   0 &   53.071709 &   -27.843693 &   2.78 &   0.15 &   5.55 &   2.43    &   0.09 &   0.02 \cr 
31 & 1.95 (1.89--2.00)  &   0 &   53.077377 &   -27.859612 &   2.54 &   0.43 &   7.04 &   2.70    &   0.07 &   0.02 \cr 
32 & 2.75 (2.57--2.78)  &   0 &   53.049751 &   -27.770971 &   2.56 &   0.16 &   9.65 &   2.68    &   0.12 &   0.02 \cr 
33 & 1.58 (1.53--1.61)  &   0 &   53.072708 &   -27.834278 &   2.49 &   0.23 &   12.96 &   2.27    &   0.08 &   0.03 \cr 
34 & 1.95 (1.86--1.97)  &   0 &   53.090752 &   -27.782473 &   2.47 &   0.21 &   4.50 &   2.01    &   0.01 &   0.03 \cr 
35 & 1.612 &   7 &   53.091747 &   -27.712166 &   2.47 &   0.13 &   12.53 &   3.32    &   0.10 &   0.02 \cr 
36 & 2.37 (2.28--2.42)  &   0 &   53.086586 &   -27.810249 &   2.41 &   0.25 &   8.90 &   1.84    &   0.04 &   0.02 \cr 
37 & 2.96 (2.87--3.05)  &   0 &   53.146378 &   -27.888807 &   2.35 &   0.28 &   4.36 &   2.92    &   0.10 &   0.03 \cr 
38 & 2.31 (2.26--2.38)  &   0 &   53.092335 &   -27.803223 &   2.50 &   0.10 &   7.60 &   1.77    &   0.12 &   0.02 \cr 
39 & 3.04 (2.99--3.21)  &   0 &   53.124332 &   -27.882696 &   2.26 &   0.18 &   10.42 &   2.71    &   0.04 &   0.03 \cr 
40 & 2.224 &   3 &   53.131123 &   -27.773195 &   2.26 &   0.17 &   10.09 &   2.17    &   0.12 &   0.03 \cr 
41 & [4.13 (3.45--4.46)]  &   0 &   53.172832 &   -27.858860 &   2.25 &   0.18 &   1.99 &   2.65    &   0.17 &   0.02 \cr 
42 & 2.34 (2.30--2.42)  &   0 &   53.091629 &   -27.853390 &   2.25 &   0.18 &   11.34 &   2.64    &   0.12 &   0.03 \cr 
43 & 2.39 (2.32--2.62)  &   0 &   53.068874 &   -27.879723 &   2.23 &   0.41 &   14.33 &   3.03    &   0.16 &   0.03 \cr 
44 &   \nodata  &    \nodata &   53.087166 &   -27.840195 &   2.21 &   0.12 &   10.59 &   2.22    &   0.18 &   0.03 \cr 
45 & [7.62 (7.15--7.93)]  &   0 &   53.041084 &   -27.837721 &   2.43 &   0.21 &   7.25 &   2.80    &   0.14 &   0.02 \cr 
46 & 1.613 &   1 &   53.104912 &   -27.705305 &   2.29 &   0.11 &   4.93 &   3.74    &   0.00 &   0.00 \cr 
47 & 2.19 (2.12--2.22)  &   0 &   53.163540 &   -27.890556 &   2.05 &   0.15 &   1.06 &   3.65    &   0.07 &   0.02 \cr 
48 & 2.543 &   6 &   53.160664 &   -27.776251 &   2.04 &   0.36 &   16.44 &   2.18    &   0.00 &   0.00 \cr 
49 & 1.87 (1.84--1.93)  &   0 &   53.053669 &   -27.869278 &   1.98 &   0.23 &   10.56 &   3.08    &   0.07 &   0.03 \cr 
50 & 1.69 (1.64--1.70)  &   0 &   53.089542 &   -27.711666 &   1.97 &   0.45 &   9.90 &   3.44    &   0.07 &   0.03 \cr 
51 & 2.32 (2.29--2.43)  &   0 &   53.067833 &   -27.728889 &   1.94 &   0.22 &   8.00 &   2.95    &   0.10 &   0.02 \cr 
52 & [4.78 (4.35--5.10)]  &   0 &   53.064793 &   -27.862638 &   1.88 &   0.24 &   0.99 &   2.73    &   0.14 &   0.03 \cr 
53 & 1.56 (1.50--1.60)  &   0 &   53.198875 &   -27.843945 &   1.86 &   0.32 &   7.09 &   3.23    &   0.00 &   0.00 \cr 
54 & [9.42 (9.35--9.83)]  &   0 &   53.181995 &   -27.814196 &   1.82 &   0.30 &   6.84 &   2.69    &   0.06 &   0.02 \cr 
55 &   \nodata  &    \nodata &   53.048378 &   -27.770306 &   1.79 &   0.15 &   6.42 &   2.74    &   0.06 &   0.02 \cr 
56 & 2.299 &   3 &   53.107044 &   -27.718334 &   1.61 &   0.25 &   5.58 &   2.75    &   0.10 &   0.03 \cr 
57 & 3.08 (3.00--3.68)  &   0 &   53.033127 &   -27.816778 &   1.72 &   0.26 &   0.48 &   2.98    &   0.00 &   0.00 \cr 
58 & [4.73 (4.39--4.90)]  &   0 &   53.183666 &   -27.836500 &   1.72 &   0.31 &   -1.2 &   2.74    &   0.19 &   0.03 \cr 
59 & 2.325 &   3 &   53.094044 &   -27.804195 &   1.84 &   0.13 &   3.58 &   1.75    &   0.07 &   0.03 \cr 
60 & 2.53 (2.41--2.60)  &   0 &   53.124584 &   -27.893305 &   1.61 &   0.25 &   6.81 &   2.88    &   0.00 &   0.00 \cr 
61 & [4.67 (4.48--5.23)]  &   0 &   53.132751 &   -27.720278 &   1.61 &   0.25 &   1.89 &   2.75    &   0.00 &   0.00 \cr 
62 & 2.94 (2.88--3.03)  &   0 &   53.080669 &   -27.720861 &   1.59 &   0.17 &   8.08 &   3.11    &   0.00 &   0.00 \cr 
63 & 1.83 (1.78--1.88)  &   0 &   53.120041 &   -27.808277 &   1.57 &   0.26 &   9.77 &   1.68    &   0.00 &   0.00 \cr 
64 & 3.26 (3.20--3.40)  &   0 &   53.117085 &   -27.874918 &   1.53 &   0.31 &   6.84 &   2.72    &   0.00 &   0.00 \cr 
65 & 1.58 (1.56--1.62)  &   0 &   53.131458 &   -27.841389 &   1.46 &   0.14 &   10.13 &   2.24    &   0.00 &   0.00 \cr 
66 & 0.653 &   1 &   53.044708 &   -27.802027 &   1.44 &   0.26 &   5.20 &   2.92    &   0.00 &   0.00 \cr 
67 & 1.69 (1.66--1.84)  &   0 &   53.072002 &   -27.819000 &   1.36 &   0.19 &   5.45 &   2.16    &   0.00 &   0.00 \cr 
68 & 5.58 &   11 &   53.120461 &   -27.742083 &   1.35 &   0.24 &   3.46 &   2.53    &   0.00 &   0.00 \cr 
69 & 2.55 (2.47--2.64)  &   0 &   53.113125 &   -27.886639 &   1.25 &   0.27 &   7.41 &   2.76    &   0.00 &   0.00 \cr 
70 & 3.14 (3.06--3.34)  &   0 &   53.141251 &   -27.872860 &   1.18 &   0.25 &   8.21 &   2.61    &   0.04 &   0.02 \cr 
71 & 1.71 (1.63--1.72)  &   0 &   53.056873 &   -27.798389 &   1.16 &   0.30 &   6.32 &   2.70    &   0.00 &   0.00 \cr 
72 & [3.76 (3.47--4.31)]  &   0 &   53.119957 &   -27.743137 &   1.11 &   0.29 &   -0.3 &   2.54    &   0.00 &   0.00 \cr 
73 & 2.19 (2.09--2.22)  &   0 &   53.142872 &   -27.874084 &   1.07 &   0.17 &   8.15 &   2.61    &   0.01 &   0.02 \cr 
74 & 0.732 &   1 &   53.093666 &   -27.826445 &   0.93 &   0.23 &   5.81 &   1.86    &   0.04 &   0.03 \cr 
75 & \nodata &   \nodata &   53.074837 &   -27.787111 &   0.84 &   0.14 &   3.49 &   2.25    &   0.00 &   0.00 \cr 
\enddata
\tablecomments{
The columns are (1) ALMA number from \citet{cowie18}, (2) redshift (three digits after the decimal point for
spectroscopic redshifts---except for the JWST NIRSpec redshift for source 68 from \citet{oesch23}---and 
two digits after the decimal point for photometric 
redshifts, plus 68\% confidence ranges from \citealt{straatman16} for photometric redshifts given in parentheses),
(3) reference for redshift
(see details below), (4) and (5) total ALMA \afluxc\ flux and error from \citet{cowie18},
(6) and (7) measured peak SCUBA-2 \afluxa\ flux and error from this work, and
(8) and (9) measured peak ALMA 2~mm flux and error from this work.
In Column~(3), `0' indicates a photometric redshift from \citet{straatman16}
(poor quality flag $Q>3$ estimates are in square brackets to distinguish them from the 
more reliable $Q<3$ estimates), while all other numbers are spectroscopic redshifts:
`1' indicates the redshift is from our own Keck DEIMOS observations, 
`2' is from K20 \citep{mignoli05}, 
`3' is from MOSDEF \citep{kriek15}, 
`4' is from \citet{casey11}, 
`5' is from \citet{szokoly04}, 
`6' is from \citet{inami17}, 
`7' is from \citet{kurk13}, 
`8' is from \citet{franco18} (B. Mobasher 2018, priv. comm.),
`9' is a CO redshift from \citet{gonzalez19}, `10' is a CO redshift from Table~\ref{tabALMAz},
and `11' is a JWST NIRSpec redshift from \citet{oesch23}.
}
\end{deluxetable*}

\subsection{Redshifts}
\label{redshifts}
There are now spectroscopic redshifts for 20 sources (see Table~\ref{tabCAT}).
Most of these were listed in Table~5 of \citet{cowie18}, but there are
five new redshifts from the present ALMA linescans (see
F.~Bauer et al.\ 2023, in preparation). We 
summarize these in Table~\ref{tabALMAz}, along with the molecular lines that 
were used to determine them.
Two others come from the ALMA SPECtroscopic Survey in the 
Hubble Ultra Deep Field \citep[ASPECS;][]{gonzalez19}.
A final new redshift is from the JWST NIRSpec survey 
First Reionization Epoch Spectroscopically Complete Observations (FRESCO;
\citealt{oesch23}).

In Table~\ref{tabCAT}, we also provide photometric redshifts 
and their 68\% confidence ranges from \cite{straatman16}, 
who used the Easy and Accurate Zphot from Yale (EAZY) code \citep{brammer08} 
to fit the FourStar Galaxy Evolution Survey (ZFOURGE) catalog from 0.3 to 8~$\mu$m. 
We note that some of these are tagged with a poor quality 
flag from EAZY, reflecting the unusual SEDs of these
high-redshift DSFGs and the limited number of band detections.
Following \citet{cowie18}, we indicate the 11 sources with 
poor quality flag $Q>3$ estimates by putting their 
photometric redshifts in square brackets.

\startlongtable
\begin{deluxetable}{ccccc}
\renewcommand\baselinestretch{1.0}
\tablewidth{0pt}
\setcounter{table}{1}
\tablecaption{ALMA Redshifts \label{tabALMAz}}
\scriptsize
\tablehead{ \\
No.  &  $z$  &    Line   &  Obs. Freq.  &  Redshift \\
& & & (GHz) & \\
(1) & (2) & (3) & (4) & (5)}
\startdata
1  & 2.5740 & & & \cr
& & $^{12}$CO($3\rightarrow2$)  & 96.763    & 2.574 \cr
& & $^{12}$CO($5\rightarrow4$)  & 161.223  & 2.574 \cr
& & $^{12}$CO($5\rightarrow4$)  & 161.516  & 2.568 \cr
& & $^{12}$CO($5\rightarrow4$)  & 159.883  & 2.604? \cr
& & $^{12}$CO($5\rightarrow4$)  & 161.399  & 2.570 \cr
2 & 3.694 & & & \cr
& & $^{12}$CO($4\rightarrow3$)  & 98.181 &  3.696 \cr
& & $^{12}$CO($6\rightarrow5$) & 147.300 &  3.694 \cr
& & $^{12}$CO($6\rightarrow5$) & 147.476 &  3.689 \cr
& & H$_2$O($2_{1,1}\rightarrow2_{0,2})$ & 160.243 &  3.693 \cr
3 &  2.648 & & & \cr
& & $^{12}$CO($3\rightarrow2$)  &  94.778  & 2.648 \cr
& & $^{12}$CO($5\rightarrow4$)  & 157.977  & 2.648 \cr
& & $^{12}$CO($5\rightarrow4$)  & 157.856  & 2.651 \cr
7 &  3.672 & & & \cr
& & $^{12}$CO($4\rightarrow3$)  &  98.681  &  3.672 \cr
& & $^{12}$CO($6\rightarrow5$)  & 148.019  & 3.672 \cr
& & $^{12}$CO($6\rightarrow5$)  & 147.980  &  3.673  \cr      
9  & 2.322  & & & \cr
& & [CI]($1\rightarrow0$)  & 148.006  &  2.325 \cr
& & HCO$^+(6\rightarrow5$) & 161.187  & 2.320 \cr
\enddata
\tablecomments{
The columns are (1) source number, (2) adopted redshift, (3) molecular line,
(4) observed frequency, and (5) redshift obtained from each molecular line.
}
\end{deluxetable}

\subsection{SCUBA-2 \afluxa\ Observations}
We have been obtaining SCUBA-2 observations of the GOODS-S for a number
of years (see \citealt{cowie18} and \citealt{barger22} for recent analyses). Our 
primary goal is to obtain the deepest possible \afluxa\ observations. In order 
to maximize the depth in the central region, we use the \textsc{CV Daisy} 
(where CV means constant speed) scan pattern. 
We choose to restrict to a radius of $5\farcm5$, where the noise is twice the central noise.
In addition, to find brighter but rarer sources in the outer regions, we use the 
\textsc{PONG-900} (where 900 refers to a $15'$ radius)
scan pattern. 
We choose to restrict to a radius of $10\farcm5$, where, again, the noise is twice the central noise.
The \textsc{CV Daisy}
scan pattern maximizes the exposure time in the center of the image, while
the \textsc{PONG-900}
scan pattern gives a wider and more uniform field coverage.
More detailed information about the SCUBA-2 scan patterns can be found in \cite{holland13}.

\begin{deluxetable}{cccc}
\tablecaption{SCUBA-2 Observations \label{tabSCUBA2}}
\setcounter{table}{2}
\tablehead{ \\
Field & Weather & Scan & Exposure\\ 
& Band & Pattern & (Hr)}
\startdata
GOODS-S   &   1   &   CV Daisy    &   59.2 \\
        &   1   &   PONG-900    &   16.2   \\
        &   2   &   CV Daisy    &   37.8  \\
        &   2   &   PONG-900    &   8.7  \\
\enddata
\end{deluxetable}

In Table~\ref{tabSCUBA2},
we summarize the total weather band~1 ($\tau_{\rm 225~GHz}<0.05$) and
weather band~2 ($0.05<\tau_{\rm 225~GHz}<0.08$) observations that we
have obtained. These are the only weather conditions 
under which \afluxa\ observations can usefully be made.
Our current \afluxa\ image has a central rms of 1.67~mJy.

We followed \citet{chen13b} for our reduction procedures, which
we describe in detail in \cite{cowie17}. 
We expect the galaxies to appear as unresolved sources at the
$7\farcs5$ resolution of the James Clerk Maxwell Telescope (JCMT)~15~m \
at \afluxa; thus, we applied a matched filter to our maps, which
provides a maximum likelihood estimate of the source strength for
unresolved sources \citep[e.g.,][]{serjeant03}.
Each matched-filter image has a PSF with
a Mexican hat shape and a FWHM corresponding to the 
telescope resolution.

Working down the ALMA \afluxc\ source
catalog from bright to faint source flux densities,
we employed an iterative procedure to extract the \afluxa\ flux densities
(positive or negative) and statistical uncertainties for each source. Given the $7\farcs5$
resolution of the \afluxa\ data, we do not need sophisticated extraction codes, 
such as those developed for the much poorer resolution BLAST ($60''$ FWHM)
or Herschel SPIRE ($35''$ FWHM) data at 500~$\mu$m \citep[e.g.,][]{bethermin10,hurley17}.
After we made each \afluxa\ flux density measurement, we removed the source from the 
\afluxa\ image. The reason for this iterative process is to remove contamination 
by brighter sources before we identify fainter sources and measure their flux densities.
However, again, given the resolution, this is not critical.

We give the SCUBA-2 \afluxa\ flux densities and statistical uncertainties in Table~\ref{tabCAT}.

\section{2~mm to \afluxc\ Flux Density Ratio Dependence on \afluxc\ Flux Density}
\label{secdep}
\begin{figure}
\includegraphics[width=8.5cm,angle=0]{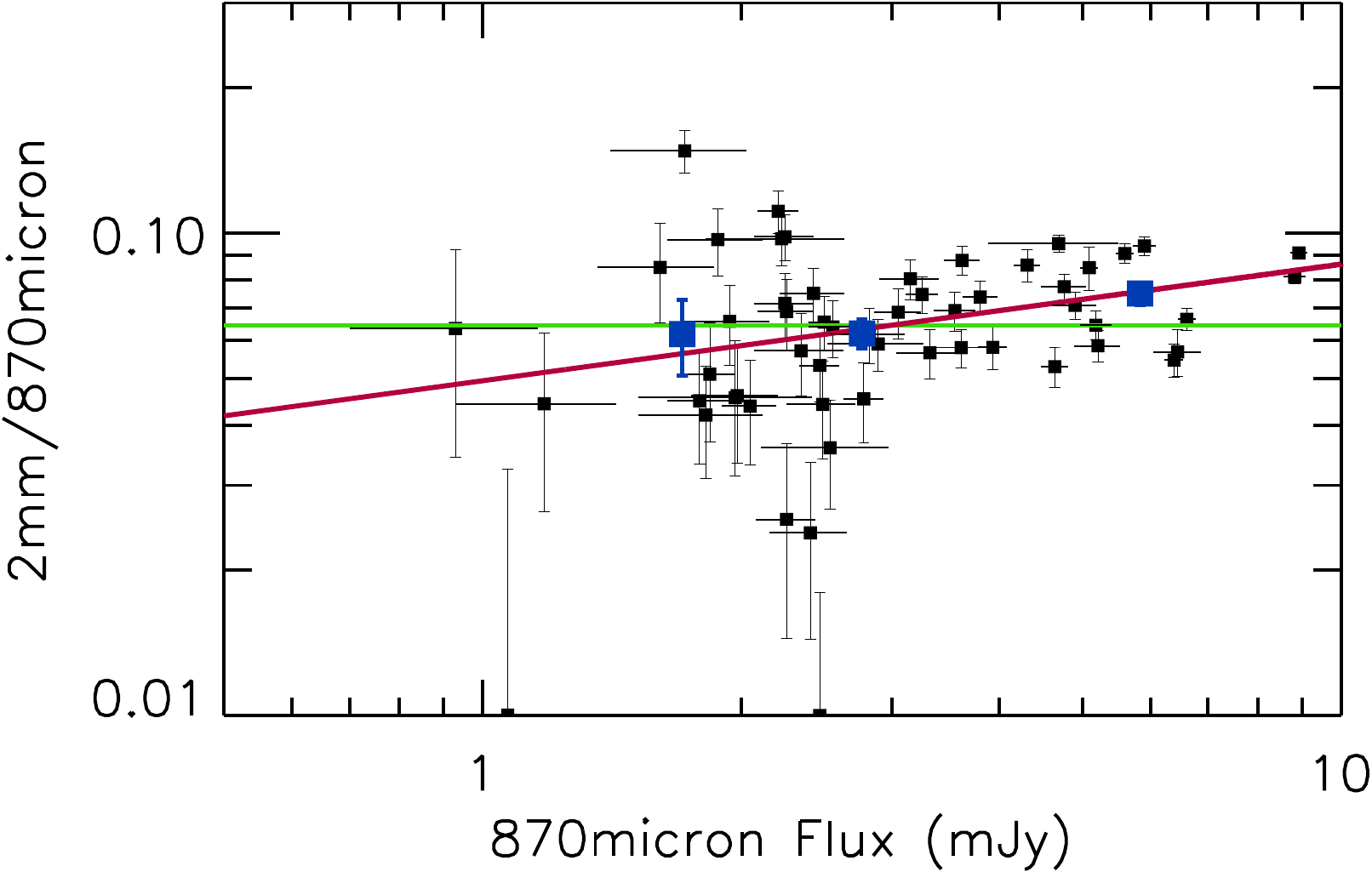}
\caption{
2~mm to \afluxc\ flux density ratio vs. \afluxc\ flux density
for every source in Table~\ref{tabCAT} that was observed at 2~mm
(black squares). The large blue squares show the 
means and errors in the means
in various flux ranges (see Table~\ref{tabMEANS}).
The green line shows the median ratio, while the red line 
shows a power law fit.
}
\label{figratiofit}
\end{figure}

In Figure~\ref{figratiofit}, we show the 2~mm to \afluxc\ flux density ratio
versus the \afluxc\ flux density
for every source in Table~\ref{tabCAT} 
that was observed at 2~mm (black squares). 
The median value of the ratio for these 55 sources is 0.064 (green line),
while the mean value is 0.066 with an error of 0.003.
We can see there is a weak positive correlation in this log-log plot.
A bisector power law fit of $\log$(\mmc) versus $\log$(\fafluxc) gives
(red line)
\begin{equation}
{{f_{2\,{\rm mm}}} \over {f_{870\,\mu{\rm m}}}} = A (f_{870\,\mu{\rm m}})^B \\,
\label{eqrelation}
\end{equation}
with $A = 0.05\pm0.002$ and $B=0.23\pm0.09$. Using a Mann-Whitney
test to compare the sources with \fafluxc\ $>2.5$~mJy with those with \fafluxc\ $<2.5$~mJy 
gives only a 0.09 probability that the two samples are different. 
We show the means and errors in the means (large blue squares
to emphasize this point (we also list these in Table~\ref{tabMEANS}). 
We conclude that the increase as a function of \fafluxc\ is not significant.

In the following sections, we will use both the median ratio and the 
bisector power law fit to convert flux densities from \afluxc\ to 2~mm.

\begin{figure*}
\centerline{\includegraphics[width=9cm,angle=0]{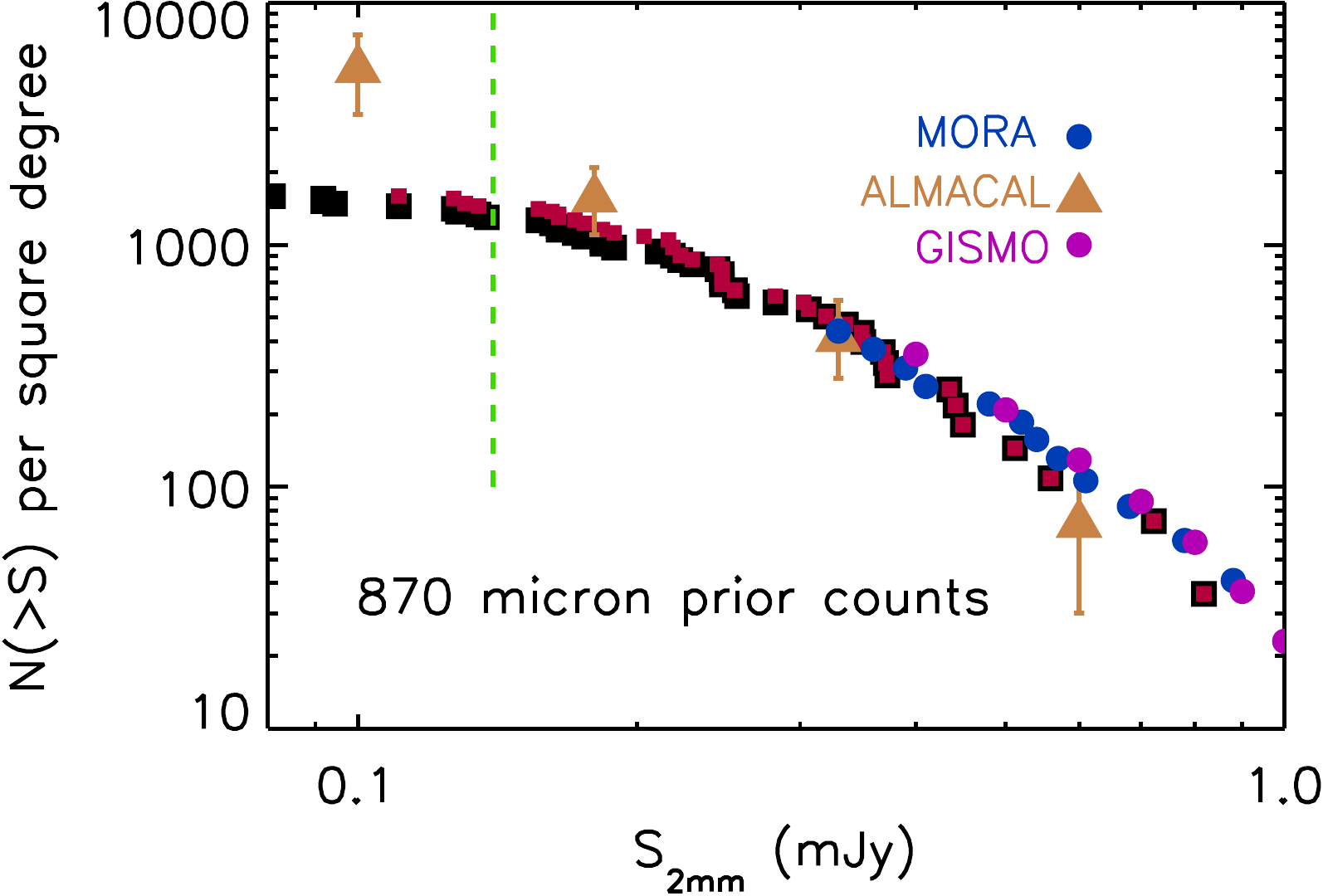}
\includegraphics[width=9cm,angle=0]{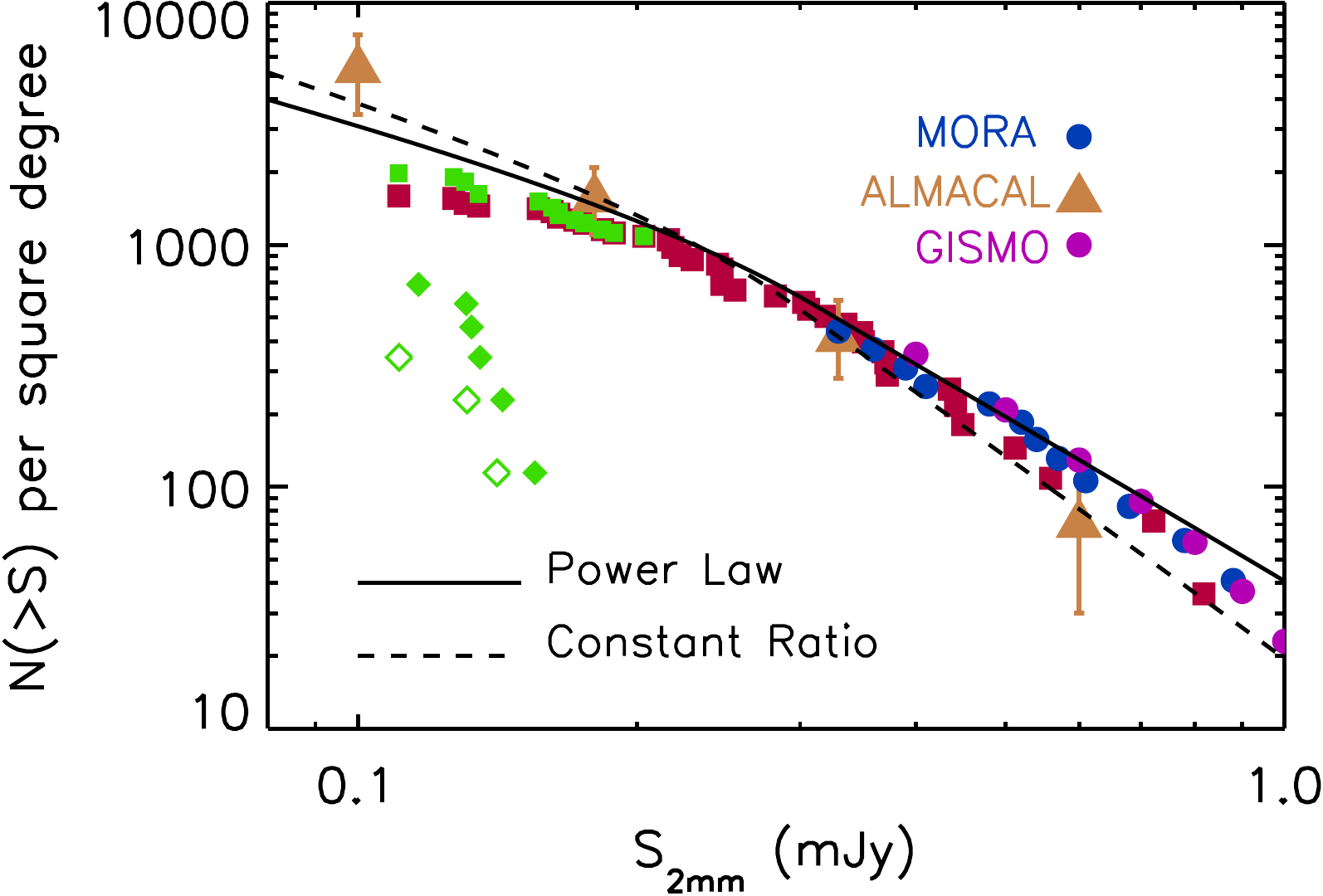}}
\caption{
Both panels show our 2~mm cumulative number counts for sources with ALMA \afluxc\ priors in the 
central 100~arcmin$^2$ of the GOODS-S. 
(a) 
The black squares show the \afluxc\ prior-based counts ($>2\sigma$ at 2~mm).
These are formally lower limits to the complete counts.
The red squares show the $>1.8$~mJy \afluxc\ prior-based counts ($>2\sigma$ at 2~mm),
corrected for the small fraction of priors without 2~mm observations.
The green dashed line shows a flux density of 0.14~mJy above which
we expect the \afluxc\ prior-based counts to be substantially complete and
below which we detect additional sources in the blank field 2~mm images (see text and (b)).
(b) The green solid diamonds show the positive blank field-based counts, and the green open diamonds
show the negative blank field-based counts.
The green solid squares show the combination of the \afluxc\ prior-based counts 
(red squares) and the positive blank field-based counts (using the 50\% weighting factor discussed in the text).
The black curves show the \afluxb\ counts from \citet{hsu16} converted to 2~mm
using either the median \mmc $=0.064$ from Section~\ref{secdep} (dashed)
or the power law in Equation~\ref{eqrelation} (solid).
In both panels, we show the IRAM GISMO (\citealt{magnelli19}; purple circles),
ALMA MORA (\citealt{zavala21}; blue circles),
and ALMACAL (\citealt{chen23}; gold triangles with error bars) 
blank field-based counts for comparison.
}
\label{fignumcts}
\end{figure*}

\begin{deluxetable}{ccccc}
\setcounter{table}{3}
\tablecaption{Means and Errors \label{tabMEANS}}
\tablehead{ \\
Number & Mean & Mean & Variance & Error in  \\
& \fafluxc & \mmc & & Mean \\
& (mJy) &  &  &}
\startdata
15   &   5.833 &   0.075 &   0.015 &  0.004 \\
28   &  2.768  &  0.062  &  0.023  & 0.004 \\
11   &   1.709  &  0.062  &  0.037  &  0.011 \\
\enddata
\end{deluxetable}

\section{Cumulative and Differential Number Counts at 2~mm}
\label{secnumcts}
In this section, we measure the cumulative and differential number counts at 2~mm.
We start with the 2~mm counts corresponding to the \afluxc\ sample.
Such counts will not be complete, since they are based on only the \afluxc\ priors.
Moreover, some of these priors do not have 2~mm observations.
We will make appropriate corrections below, but here we note that since the \afluxc\ 
sample is highly complete to near 2~mJy,
and given the median \mmc=0.064 from Section~\ref{secdep},
we expect the 2~mm counts will be near-complete to 0.13~mJy.

We take as our \afluxc\ priors the 70 sources from \citet{cowie18} that lie within 
the central 100~arcmin$^2$ of the field. Fifty-one of these have 2~mm observations.
To determine which signal-to-noise threshold to adopt---we already know it
can be lower than one would choose for a blank field selection, 
given our use of priors---we
estimate the expected level of spurious 2~mm detections by measuring
the flux densities in the 2~mm images at random positions. We find
one spurious 2~mm detection if we use a $2\sigma$ threshold, and
0.25 if we use a $2.5\sigma$ threshold.
We therefore adopt a $2\sigma$ threshold, which
49 of the 51 sources satisfy.

In Figure~\ref{fignumcts}(a), we show as black squares the cumulative
counts for the 49 \afluxc\ priors detected above a $2\sigma$ threshold at 2~mm.
These are formally lower limits to the complete number counts.
We form the counts from the number of sources above a
given flux density divided by 100~arcmin$^2$, 
the area of our \afluxc\ priors.

Next, we correct the counts to allow for the \afluxc\ priors
that do not have $>2\sigma$ 2~mm flux densities. Of the 50
priors with \afluxc\ flux densities above 1.8~mJy in the central 100~arcmin$^2$ of the field,
45 have 2~mm observations (see Section~\ref{almaobs} for why 5 sources were excluded 
from the 2~mm observations), 
only one of which is not detected above the $2\sigma$ threshold. 
We recompute the cumulative number counts for this sample of 50 sources, 
assigning 2~mm flux densities to the 6 missing sources using their \afluxc\ 
flux densities and the median \mmc\ $=0.064$.
We show these counts as the red squares in
Figure~\ref{fignumcts} (both panels), and we hereafter refer to them as our
prior-based 2~mm counts.
The correction is small and only appears at the faintest flux densities

In Figure~\ref{fignumcts} (both panels),
we compare our prior-based 2~mm counts with blank field-based
2~mm counts from the literature. The brighter counts come
from the IRAM GISMO sample of \citet{magnelli19} (purple circles)
and the ALMA MORA sample of \citet{zavala21} (blue circles).
We do not show the GISMO sample of \citet{staguhn14}, whose
counts are high compared with the other literature counts.
Note that all of these samples contain a relatively small
number of sources ($\lesssim15$ sources in each).
The deeper counts are from the ALMACAL sample of \citet{chen23} (gold
triangles). Above a 2~mm flux of $0.2$~mJy, our prior-based 2~mm counts agree
well with the literature blank field-based 2~mm counts.

In addition to the 2~mm sources found by using our \afluxc\ priors, we may look at the
blank field images to see if there are any sources not found by using our priors.
To do so, we restrict each 2~mm ALMA image to a radius of $28''$ for a relatively uniform 
rms noise of $<0.027$~mJy.
Then, after masking the targeted \afluxc\ ALMA sources,
our 2~mm images provide deep, blank field 
observations over an area of 32~arcmin$^2$. 
Given the large number of independent beams (just over 50000)
in the area, we adopt a fairly high selection threshold
of $4.5\sigma$. At this level, we expect about 0.4 false positive sources.

When we search the 32~arcmin$^2$ area,
we find 6 additional sources with aperture and primary-beam corrected 
2~mm flux densities between 0.12 and 0.16~mJy.
(Seven of the \afluxc\ priors have 2~mm flux densities in this range, six of
which have an \afluxc\ flux above 1.8~mJy.)
We do not attempt to make any corrections for clustering that
might bias the sample \citep[e.g.,][]{bethermin20}.
In order to test our false positive estimation, we run the
same procedure on the negatives of the images.
This yields 3 sources, which suggests that as many as
50\% of the additional sources in the images may be spurious.
(If we change the selection threshold from $4.5\sigma$ to $5\sigma$, then 
we have 2 additional sources in the images and none in the negatives of
the images.)

We also try to check the reliability of the 6 additional sources in the 2~mm images
by looking to see whether any are detected at 1.2~mm or 3~mm.
However, only 2 of the 6 are well covered by the 1.2~mm observations.
Neither is significantly detected. All 6 are covered
by the 3~mm observations, but the 3~mm data
are too shallow to provide useful constraints.

In Figure~\ref{fignumcts}(b),
we show the cumulative number counts for the additional sources in the images 
(green solid diamonds) and for those in the negatives of the images (green open diamonds).
We use the 50\% weighting factor to combine 
the positive blank field-based counts with the \afluxc\ prior-based counts to obtain
a final estimate of the counts (green squares).

\begin{figure}
\includegraphics[width=8.5cm,angle=0]{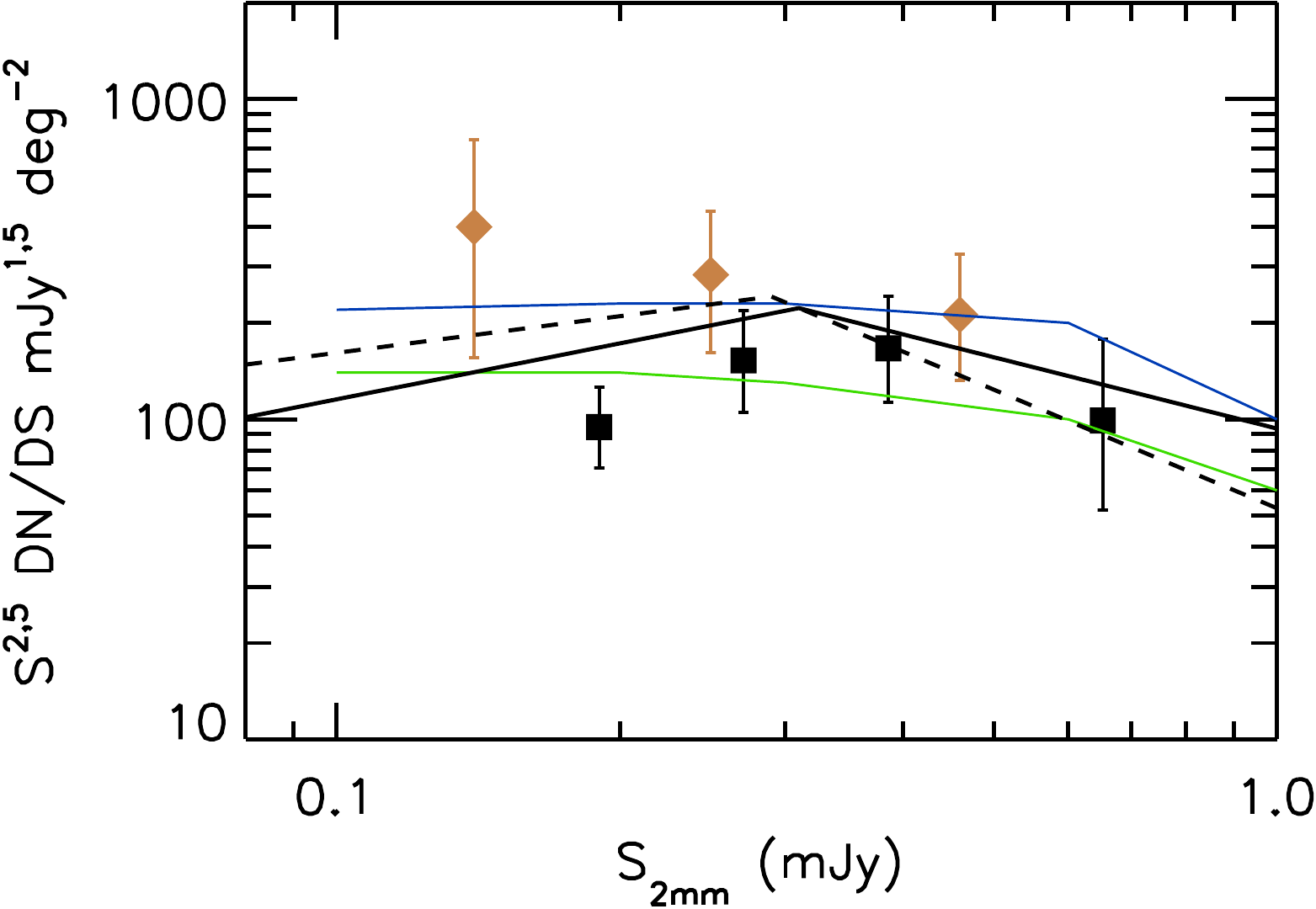}
\caption{
Differential number counts above 0.14~mJy for our corrected data
(black squares). The error bars are 68\% confidence Poisson
uncertainties. The flux density bins are 0.14--0.23, 0.23--0.32,
0.32--0.5, and 0.5--1~mJy, which
contain 15, 10, 9, and 4 sources, respectively.
The ALMACAL counts \citep{chen23}
are shown with gold diamonds. Their faintest bin contains only 4 sources.
The black curves show the \afluxb\ counts from \citet{hsu16} converted to 2~mm
using either the median \mmc $=0.064$ from Section~\ref{secdep} (dashed)
or the power law in Equation~\ref{eqrelation} (solid).
The models of \citet{lagos20} and \citet{popping20} 
(taken from \citealt{chen23}) are shown with
blue and green curves, respectively.
}
\label{figdiffcts}
\end{figure}

As discussed in \citet{zavala21}, cumulative number counts are
clearer for these small samples. However,
it is easier to show errors when using differential number
counts. Thus, in Figure~\ref{figdiffcts}, we show the differential counts 
above 0.14~mJy for our corrected data (black squares).
The error bars are 68\% confidence Poisson
uncertainties on the number of sources in each bin.
We compare these with
the ALMACAL counts (gold diamonds) \citep{chen23}. 
Above 0.2~mJy, the two samples are consistent within the errors,
while below 0.2~mJy, our counts are
low compared to ALMACAL. However, their lowest
point is based on only 4 sources and hence is quite uncertain. 

We also show in Figure~\ref{figdiffcts} the models of \citet{lagos20} (blue) and
\citet{popping20} (green), as taken from \citet{chen23}. 
Above 0.2~mJy, these models are broadly
consistent with both the ALMACAL and present counts, 
while below 0.2~mJy, they lie between the two samples.

Ultradeep galaxy number counts, which nearly fully resolve the EBL, have 
now been obtained  at \afluxb\ or \afluxc\ 
\citep[e.g.,][]{hsu16,bethermin20,chen23} and 
at 1.1~mm or 1.2~mm \citep[e.g.,][]{fujimoto16,munoz18,munoz23,gonzalez20}.
Here we make comparisons with \citet{hsu16} by converting
their differential number counts, which take the form of a broken power law
\begin{equation}
\frac{dN}{dS}=\left\{
\begin{matrix}
N_0\left ( \frac{S}{S_0} \right )^{-\alpha} ~~{\rm if}~S\leq S_0 \\
N_0\left ( \frac{S}{S_0} \right )^{-\beta} ~~{\rm if}~S > S_0 \\
\end{matrix}
\right. \,,
\end{equation}
to 2~mm.
For the median \mmc=0.064 conversion, $\alpha$ is 2.12, $\beta$ is 3.73, 
$S_0$ is 0.29~mJy, and the normalization, $N_0$, is 5340~mJy$^{-1}$~deg$^{-2}$.
For the Equation~\ref{eqrelation} conversion, $\alpha$ is 1.92, $\beta$ is 3.24, 
$S_0$ is 0.31~mJy, and $N_0$ is 4160~mJy$^{-1}$~deg$^{-2}$.
We show these power laws on Figure~\ref{figdiffcts}.
Above 0.2~mJy, they match well to both the ALMACAL and present counts,
while below 0.2~mJy, they lie between the two samples. The power laws are
also broadly consistent with the \citet{lagos20} and \citet{popping20} models.

We then integrate these broken power laws to get the cumulative counts, which
we show as black curves in Figure~\ref{fignumcts}(b). These curves provide a good match
to all of the 2~mm data.

Above 0.2~mJy, where we expect the sample to be essentially complete based on 
the \afluxc\ selection (see Figure~\ref{fignumcts}),
our measured contribution to the 2~mm EBL is 
$0.63\pm0.09$~Jy~deg$^{-2}$.
Comparing with the total values of $3.0\pm5.8$~Jy~deg$^{-2}$
from \citet{odegard19} based on the Planck HFI data
and 6~Jy~deg$^{-2}$ inferred by \cite{chen23} by extrapolating the COBE FIRAS
data to 2~mm, this corresponds to 10 to 19\%. But the uncertainties on the total EBL
measurements are substantial.

While recognizing the uncertainties inherent in extrapolating beyond what is measured,
we can use the converted \citet{hsu16} counts to extrapolate the
contribution to the EBL to fainter 2~mm flux densities than measured in order
to estimate how faint future 2~mm measurements---such as those that will be
made with TolTEC---may need to be to resolve the EBL substantially.

With the median conversion, we would resolve the 2~mm EBL at $\sim0.004$~mJy,
while with the power law conversion, we would need to reach a flux
$<0.001$~mJy to resolve substantially the 2~mm EBL.
This may be difficult to achieve with TolTEC and may require ALMA observations
of lensing clusters, such as those carried out at 1.1~mm or 1.2~mm 
\citep[e.g.,][]{fujimoto16,munoz18,gonzalez20}.

We emphasize that these estimates depend  on the extrapolation of
\mmc\ well below the values at which they were measured. If there
are sources with higher values of this
ratio at lower 2~mm flux densities, then the EBL contributions
could be higher and the resolution of the
2~mm EBL could occur at higher flux densities.

\section{2~mm Based Redshift Estimates}
\label{sechighz}

\begin{figure*}
\includegraphics[width=8.5cm,angle=0]{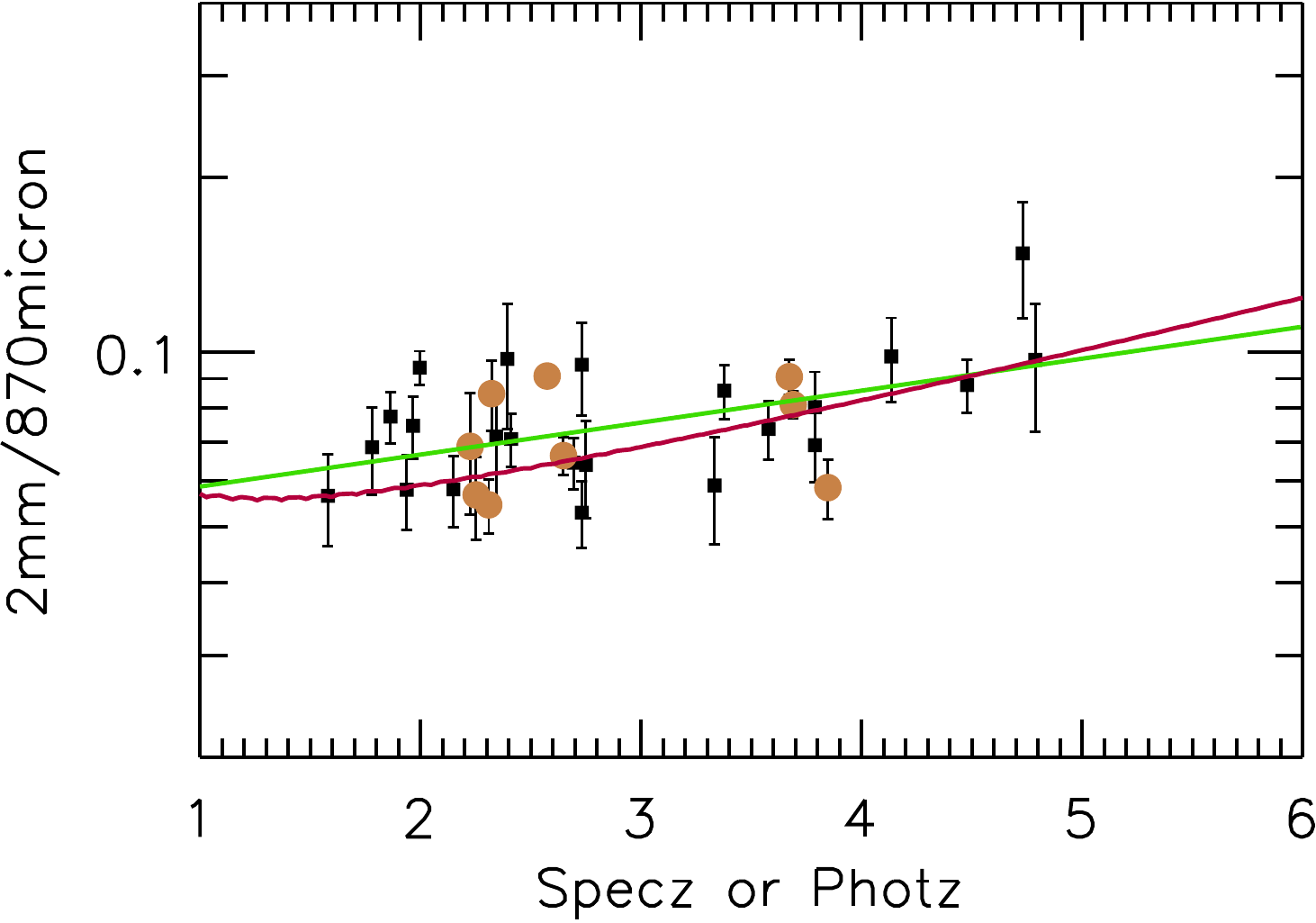}
\hskip 0.5cm
\includegraphics[width=8.75cm,angle=0]{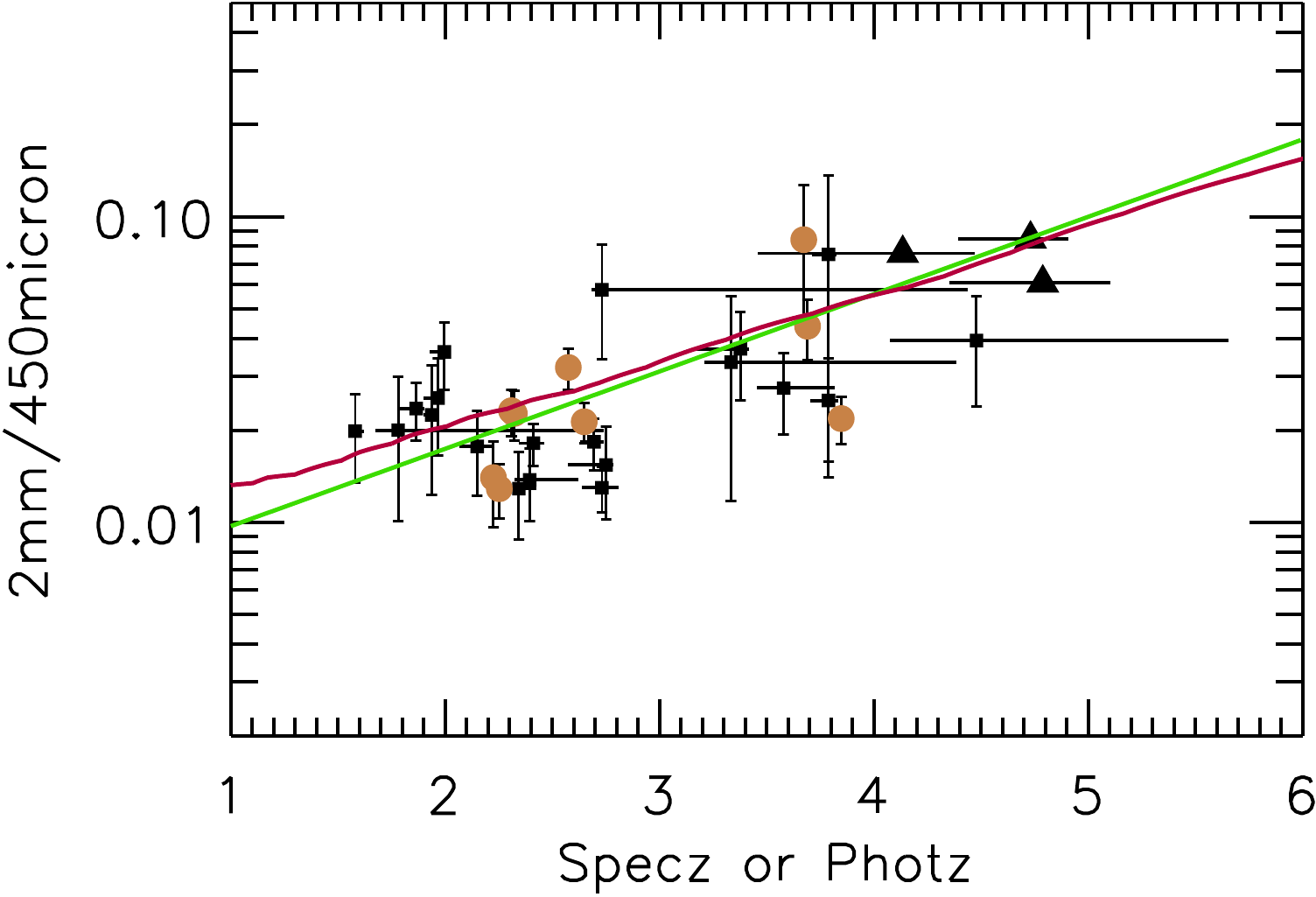}
\caption{
(a) \mmc\ and (b) \mma\ vs. spectroscopic redshift (gold circles) or 
\citet{straatman16} photometric redshift (black symbols; for clarity, we only show
the 68\% confidence ranges in (b))
for the sources in Table~\ref{tabCAT} with a corrected aperture flux density
above 0.14~mJy.
The uncertainties in (a) are dominated by the \fafluxd\ uncertainties, while
those in (b) are dominated by the \fafluxa\ uncertainties.
The upward pointing triangles in (b) show the $1\sigma$ lower limits on the flux ratios
for the three sources not detected at \afluxa.
The green curves show power law fits to the data (Equations~\ref{powerlaw} and \ref{powerlaw2}).
The red curves show the flux ratios using Arp~220 as a template
for the SED shape \citep{silva98}. 
}
\label{figratios}
\end{figure*}

As we discussed in the Introduction,
one of the primary motivations for pushing to longer
wavelengths than \afluxc\ is to detect a larger
fraction of high-redshift galaxies (see Figure~\ref{figselect}(a)).
However, this
still leaves the problem of determining the redshifts
for the detected sources, and, in particular, for the high-redshift galaxies.
Many submillimeter or millimeter detected galaxies are too faint 
for optical/NIR spectroscopic redshifts, or in a small number of cases, 
even for photometric redshifts.
For the 55 sources with measured 2~mm flux densities in Table~\ref{tabCAT},
only 13 have well-determined spectroscopic redshifts. Even including in this count
the handful of sources that were not observed at 2~mm
because they had existing spectroscopic redshifts, less than a third of all the observed sources
have spectroscopic redshifts. No source in Table~\ref{tabCAT} has a spectroscopic redshift greater than 4.
 
Ideally, one can measure redshifts using spectral observations in the 
millimeter, as we have done for some of the brighter 2~mm sources in our 
sample (F.~Bauer et al.\ 2023, in preparation), and as ASPECS has done 
for several of the slightly fainter sources \citep{gonzalez19}.
Alternatively, one can estimate redshifts by fitting to the entire
FIR SED \citep[e.g.,][]{battisti19, dud20}. 
This was done for the present sample in \cite{cowie18},
though this type of analysis does have uncertainties
due to redshift degeneracies with the temperatures
of the dust SEDs \citep[e.g.,][]{casey19,jin19,cortzen20}.

Redshifts can also be roughly
estimated using submillimeter or millimeter flux ratios, which
allow one to work with much more limited data.
Here we focus on the purely empirical relation
between measured flux ratio (from the
present \afluxa, \afluxb, and 2~mm data)
and spectroscopic or photometric redshift (estimated from optical/NIR data).
The 2~mm data improve these estimates by providing a wider
wavelength separation.

In Figure~\ref{figratios}, we plot for the sources
with a 2~mm flux density above 0.14~mJy (all of these are detected above $3\sigma$)
(a) \mmc\ and (b) \mma\ versus available spectroscopic (gold circles) 
or photometric (black) redshift. Fitting the ratios versus redshift with error-weighted power law fits
(green curves) gives, respectively,
\begin{equation}
\log(f_{2\,{\rm mm}}/f_{870\,\mu{\rm m}}) = 0.067 z - 1.400 
\label{powerlaw}
\end{equation}
and
\begin{equation}
\log(f_{2\,{\rm mm}}/f_{450\,\mu{\rm m}}) = 0.268 z - 2.319 \,.
\label{powerlaw2}
\end{equation}
Bisector fits give almost identical results, with the right-hand side of
Equation~\ref{powerlaw} becoming $0.065z-1.329$, and the
right-hand side of Equation~\ref{powerlaw2} becoming $0.278z-2.361$.

For \btradmm, the power law fit is almost identical to that given
in \cite{barger22}, namely, 
\begin{equation}
\log(f_{870\,\mu{\rm m}}/f_{450\,\mu{\rm m}})=0.16 z - 0.96 \,,
\label{powerlaw3}
\end{equation}
which was based on both this field (though the 450~$\mu$m data were not as deep)
and the GOODS-N.

Unfortunately, the dependence of \mmc\ on redshift is
too shallow, especially given the dispersion, for it to be useful in estimating redshifts.
However, \mma\ has a steeper dependence
on redshift and hence can provide better redshift estimates than
\btradmm. Both of these
estimates depend on the passage of the \afluxa\ band
through the rest-frame peak 100~$\mu$m region; thus, they are critically
dependent on the short-wavelength data.

\begin{figure}
\centerline{
\includegraphics[width=8.25cm,angle=0]{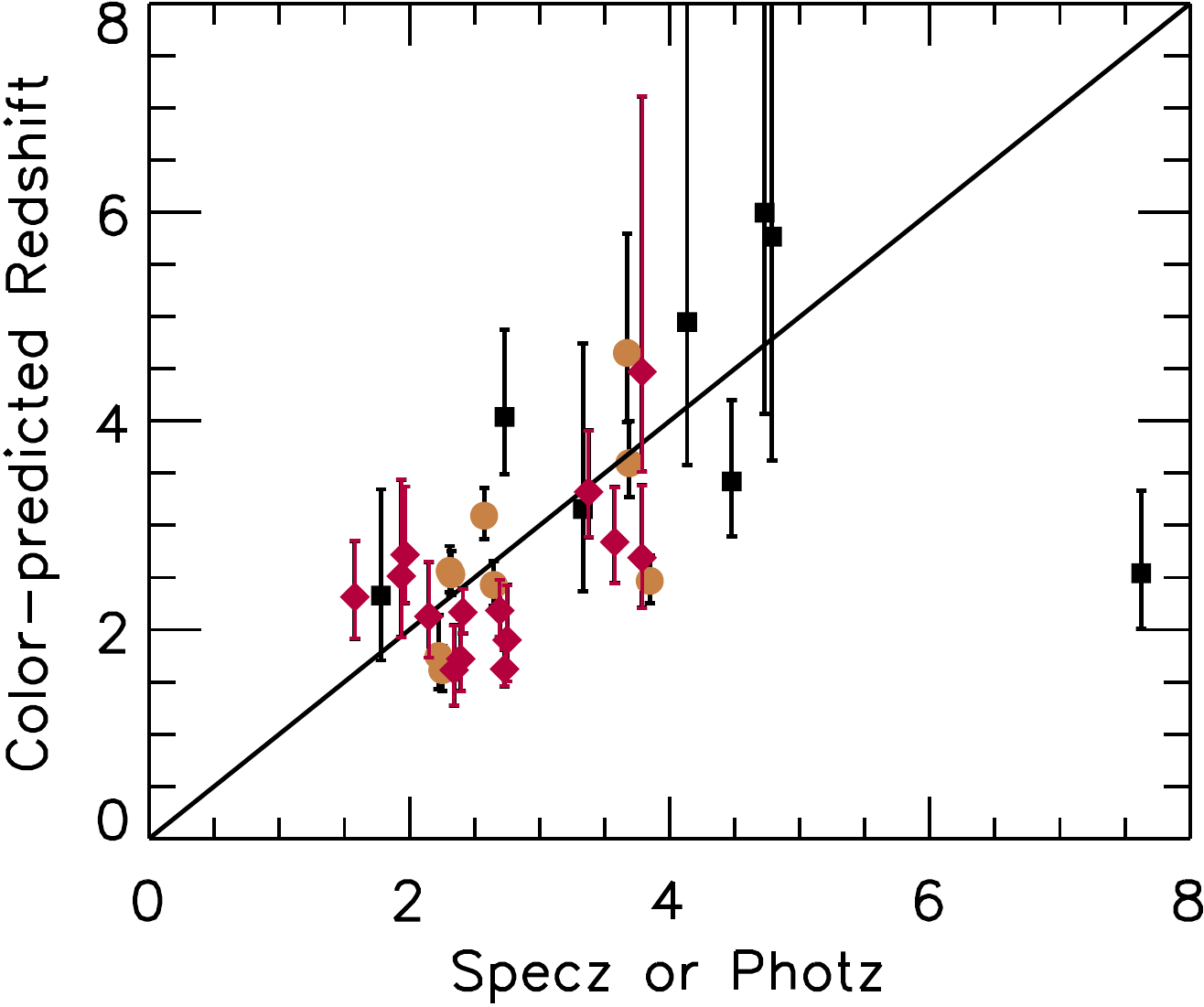}
}
\caption{
Color-predicted redshift based on \mma\
vs. spectroscopic redshift (gold circles) or \citet{straatman16} 
photometric redshift (red diamonds for robust $Q<3$, and black squares for poorer quality $Q>3$) for
the sources in Table~\ref{tabCAT} with \fafluxd\ $>0.14$~mJy.
All errors are 68\% confidence.
}
\label{z_color}
\end{figure}

As we illustrate in Figure~\ref{z_color}, the \mma\ color-predicted redshift
for the sources with a 2~mm flux density above 0.14~mJy
is correlated with the spectroscopic and photometric redshift samples,
with a similar spread for both.
We show uncertainties on the color-predicted redshifts that
correspond to the 68\% range in the flux density ratio.
Within these uncertainties, only five of the sources
(7, 26, 41, 52, and 58 in Table~\ref{tabCAT}) could
lie at $z>5$. However, source~7, which has a color-predicted 
redshift range from 3.98 to 5.80,
has a CO redshift of 3.672 (Table~\ref{tabALMAz}).

This leaves only four sources that could lie at $z>5$ based on the color-predicted redshifts.
The photometric redshifts for these four sources are between 3.8 and 4.8. Since the photometric
redshift for source~26 is robust ($Q<3$), we do not consider it further as 
a $z>5$ candidate. However, the photometric redshifts for the remaining sources are considered 
to be of poorer quality ($Q>3$), leaving us with a total of three $z>5$ candidates.

Note that only one source has a $z_{\rm phot} > 5$ (source~45):
\cite{straatman16} give $z_{\rm phot}=7.62$ ($Q>3$), while \cite{santini15} give
$z_{\rm phot}=6.62$. However, this is a poorly determined photometric redshift,
and the color-predicted redshift is more consistent with a $z<3.5$ solution 
(see Figure~\ref{z_color}).

\begin{figure}
\centerline{
\includegraphics[width=8.5cm,angle=0]{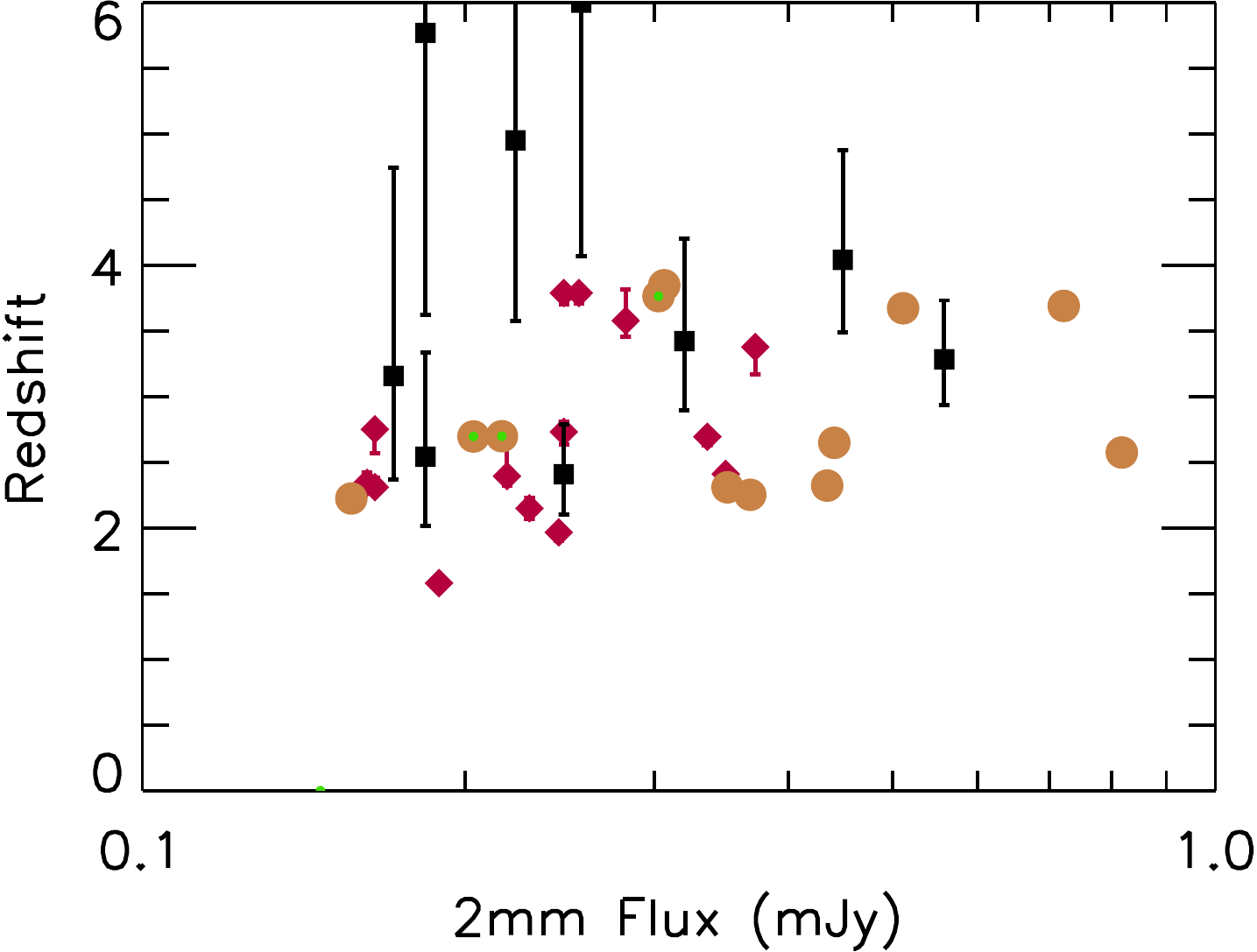}
}
\caption{
Redshift vs. \fafluxd\ for the sources in Table~\ref{tabCAT} with \fafluxd\  $>0.14$~mJy.
Black squares show color-predicted redshifts
based on \mma, while gold circles show sources with spectroscopic redshifts
and red diamonds show sources with \citet{straatman16} photometric redshifts (robust $Q<3$ only).
All errors are 68\% confidence. We have included three sources using $0.064\times$\fafluxc\ 
where \fafluxd\ was not measured (central green circles)
}
\label{z_color_flux}
\end{figure}

As we illustrate in Figure~\ref{z_color_flux},
there is no strong dependence of redshift on \fafluxd.
A bisector fit gives a weak gradient of $z= 3.38  +   0.64 \log$ \fafluxd,
with the brighter sources being at a slightly
higher redshift. However, a Mann-Whitney test comparing the \fafluxd\ $>0.33$~mJy
sample with the \fafluxd\ $=$ 0.14--0.33~mJy sample gives a $p$ value
of 0.74, so the two distributions are not
significantly different at the 0.05 level.
The bisector fit is shallower than the model in Figure~3 of
\citet{bethermin15}, but given the small range in flux density, and the uncertainties
in the slope, it could be broadly consistent.

In \cite{cowie18}, we determined star formation rates (SFRs) for the ALMA \afluxc\ 
sample using MAGPHYS \citep{dacunha15}.
In MAGPHYS, SFRs are computed for a \citet{chabrier03} initial mass function (IMF). 
However, since there
is a near-constant dependence of SFR on 2~mm flux (see Figure~\ref{figselect}(a)),
we can also simply estimate the SFR from the 2~mm flux. 
Doing a linear fit to the MAGPHYS SFRs versus the 2~mm flux
gives SFR $=1750\times$(\fafluxd\ in\,mJy) 
M$_\odot$~yr$^{-1}$, which we adopt here.

We can compare this with estimates in the literature of the relation between SFR and
\afluxb\ flux (converted to a Chabrier IMF) using \mmc\ = 0.064. These give
SFR $=1970\times$(\fafluxd\ in\,mJy) M$_\odot$~yr$^{-1}$ \citep{barger14} and
SFR $=2100\times$(\fafluxd\ in\,mJy) M$_\odot$~yr$^{-1}$ \citep{cowie17},
which are reasonably consistent with the relation in the previous paragraph,
given the uncertainties.

Numerous estimates of the dusty star formation history have been made from submillimeter
observations
\citep[e.g.,][]{barger00,barger14,chapman05,wardlow11,casey13,swinbank14,cowie17}.
However, determining the contributions of dusty galaxies at very high redshifts has been
challenging. Here we focus on whether the 2~mm data find
significant  star formation at high redshifts (here $z>5$),
and we make comparisons with the 2~mm results of \citet{zavala21} and \citet{cooper22}.

Using the SFR estimated from the 2~mm flux and the spectroscopic,
photometric $(Q<3)$,
or else \mma\ color-predicted redshift (see Figure~\ref{z_color_flux}),
we can estimate the star formation history.
The 2~mm flux limit of 0.14~mJy corresponds to a SFR of 250~M$_\odot$~yr$^{-1}$. 
Above this limit, we find the SFR densities given in Table~\ref{tabSFRD}.
These values are considerably
higher than those given in \citet{cooper22} due to the depth of our observations.
We find a peak in the $z=2$--3 range and a rapid drop-off to the highest
redshifts. 

Even if we place all three $z>5$ candidates at these
redshifts, they only contain about 6\% of the star formation of all the DSFGs
observed in the area (1200~M$_\odot$~yr$^{-1}$ versus 19000~M$_\odot$~yr$^{-1}$).
This is a smaller contribution than claimed by \citet{zavala21} at such high
redshifts ($\sim35$\% at $z=5$ and 20--25\% in the redshift range $z=$ 6--7).
We conclude that only a very small fraction of DSFG formation occurs at high redshift.

\begin{deluxetable}{ccc}
\setcounter{table}{4}
\tablecaption{SFR Densities \label{tabSFRD}}
\tablehead{ \\
Redshift & SFR Density & 68\% Confidence \\
& \multicolumn{2}{c}{(M$_\odot$~yr$^{-1}$~Mpc$^{-3}$)}
}
\startdata
1.2--2  &   0.003 &  0.001--0.007  \\
2--3   &  0.028  &  0.022--0.036  \\
3--4   &   0.023  &  0.016--0.032  \\
4--5  &  0.004 &  0.001--0.010  \\
5--6  & 0.003 & 0.001--0.007 \\
\enddata
\end{deluxetable}

\section{Summary}
\label{secsum}

Starting from an ALMA \afluxc\ sample in the GOODS-S obtained
from targeted observations of SCUBA-2 \afluxb\ sources,
we made ALMA spectral line scans that provide 2~mm continuum observations
down to below a 2~mm flux density of 0.1~mJy.
We also deepened our SCUBA-2 \afluxa\ observations of the field.
Our primary results are as follows.

(1) Using every source that we observed at 2~mm,
we found a median \mmc\ of 0.064 (the mean ratio is 0.066 with an error of 0.003
and determined a bisector power law fit to $\log$(\mmc) versus $\log$(\fafluxc).
We used these to convert \afluxc\ flux densities to 2~mm.

(2) We measured the cumulative and differential number counts at 2~mm using
the \afluxc\ sources as priors. We corrected the counts for the small number of sources 
without $>2\sigma$ 2~mm flux densities. We also searched for additional 2~mm sources 
not found by our priors. After estimating the number of false positives by searching
for sources in the negatives of the images, we corrected the counts.
We determined that our cumulative and differential 2~mm number counts
are in reasonable agreement with the literature above 0.2~mJy.

(3) Above 0.2~mJy, we measured a contribution to the 2~mm EBL of $0.63\pm 0.09$~Jy~deg$^{-2}$,
which corresponds to 10 to 19\% of the total EBL given by
\citet{odegard19} and \citet{chen23}, respectively,
but the uncertainties on these values are substantial.
While recognizing the uncertainties inherent in
extrapolating beyond what is measured, we estimated that 
in order to resolve the 2~mm EBL substantially,
2~mm flux density measurements below 
0.004~mJy---and possibly as faint as 0.001~mJy---will need to be reached.

(4) We determined that the 2~mm to \afluxa\ flux density ratio provides an
estimate of the spectroscopic and photometric redshifts of the sources.
We found no significant dependence of the redshifts on
the 2~mm flux density for sources with 2~mm flux densities between 0.14
and 1~mJy.

(5) We only identified three galaxies that may lie at $z>5$.
Our observations measure galaxies with SFRs in excess of 
250~M$_\odot$~yr$^{-1}$. For these galaxies, the SFR densities 
fall by a factor of $\sim9$ from $z=2$--3 to $z=5$--6.

\bigskip
\section*{Acknowledgements}
We thank the anonymous referee for constructive comments that helped us to improve 
the manuscript. We gratefully acknowledge support for this research from 
NASA grant 80NSSC22K0483 (L.~L.~C.), 
a Kellett Mid-Career Award and a WARF Named Professorship from the 
University of Wisconsin-Madison Office of the 
Vice Chancellor for Research and Graduate Education with funding from the 
Wisconsin Alumni Research Foundation (A.~J.~B.),
the Millennium Science Initiative Program -- ICN12\_009 (F.~E.~B), 
CATA-Basal -- FB210003 (F.~E.~B), and FONDECYT Regular -- 1190818 (F.~E.~B) 
and 1200495 (F.~E.~B.).

The National Radio Astronomy Observatory is a facility of the National Science
Foundation operated under cooperative agreement by Associated Universities, Inc.
This paper makes use of the following ALMA data: 
ADS/JAO.ALMA\#2021.1.00024.S. \\ 
ALMA is a partnership of ESO (representing its member states), NSF (USA), and NINS (Japan), 
together with NRC (Canada), MOST and ASIAA (Taiwan), and KASI (Republic of Korea),
 in cooperation with the Republic of Chile. The Joint ALMA Observatory is operated by 
 ESO, AUI/NRAO, and NAOJ.
 
 The James Clerk Maxwell Telescope is operated by the East Asian Observatory on 
behalf of The National Astronomical Observatory of Japan, Academia Sinica Institute 
of Astronomy and Astrophysics, the Korea Astronomy and Space Science Institute, 
the National Astronomical Observatories of China and the Chinese Academy of 
Sciences (grant No. XDB09000000), with additional funding support from the Science 
and Technology Facilities Council of the United Kingdom and participating universities 
in the United Kingdom and Canada. 

We wish to recognize and acknowledge 
the very significant cultural role and reverence that the summit of Maunakea has always 
had within the indigenous Hawaiian community. We are most fortunate to have the 
opportunity to conduct observations from this mountain.

\facilities{ALMA, JCMT}

\software{\textsc{casa} \citep{mcmullin07}}

\bibliography{smmref}

\end{document}